\documentclass[onefignum,onetabnum]{siamart190516}

\usepackage{amssymb}
\usepackage[bbgreekl]{mathbbol}
\usepackage{mathrsfs}
\usepackage{graphicx,epstopdf}
\usepackage{amsfonts}
\usepackage{algorithm, algorithmic}
\usepackage{multirow}
\usepackage{subcaption}
\usepackage{setspace}
\usepackage{booktabs}
\usepackage{ulem}

\newsiamthm{remark}{Remark}

\newcommand{\Veff}{V_{\rm{eff}}}
\newcommand{\Vps}{v_{\rm{ps}}}
\newcommand{\VH}{v_{\rm{H}}}
\newcommand{\Vxc}{v_{\rm{xc}}}
\newcommand{\Ne}{N_{\rm e}}
\newcommand{\Nb}{N_{\rm b}}
\newcommand{\Nn}{N_{\rm atom}}
\newcommand{\R}{\mathbb{R}}
\newcommand{\LL}{\mathbb{L}}
\newcommand{\Z}{\mathbb{Z}}
\newcommand{\G}{\pmb{G}}
\newcommand{\rr}{\pmb{r}}
\newcommand{\Ec}{E_{\rm c}}
\newcommand{\EcA}{E_{\rm c,A}}
\newcommand{\EcB}{E_{\rm c,B}}
\newcommand{\Ecopt}{E_{\rm c}^{\rm opt}}
\newcommand{\Eg}{E_{\rm g}}
\newcommand{\A}{\textit{A}}
\newcommand{\dd}{~{\rm d}}
\newcommand{\res}{{\rm Res}}

\newcommand{\norma}[1]{\|#1\|_a}
\newcommand{\normd}[1]{\|#1\|_{a'}}
\newcommand{\uf}[1]{\textrm{#1}}
\newcommand{\vect}[1]{\boldsymbol{#1}}

\title{An adaptive planewave method for electronic structure calculations}

\author{Beilei Liu\thanks{School of Mathematical Sciences, Beijing Normal University, No.19, Xinjiekouwai Street, Beijing 100875, China (\email{beilei@mail.bnu.edu.cn}, \email{chen.huajie@bnu.edu.cn}).
}
\and Huajie Chen\footnotemark[2]
\and Genevi\`{e}ve Dusson\thanks{Laboratoire de Math\'ematiques de Besan\c{c}on, UMR CNRS 6623, Universit\'e Bourgogne Franche-Comt\'e, 16 route de Gray, 25030 Besan\c{c}on, France (\email{genevieve.dusson@math.cnrs.fr}).
}
\and Jun Fang\thanks{Institute of Applied Physics and Computational Mathematics, Fenghao East Road 2, Beijing 100094, China (\email{fang\_jun@iapcm.ac.cn}).
}
\and Xingyu Gao\thanks{Laboratory of Computational Physics, Institute of Applied Physics and Computational Mathematics, Huayuan Road 6, Beijing 100088, China (\email{gao\_xingyu@iapcm.ac.cn}).
}}

\begin{document}

\maketitle

\begin{abstract}
  We propose an adaptive planewave method for eigenvalue problems in electronic structure calculations. The method combines {\it a priori} convergence rates and accurate {\it a posteriori} error estimates into an effective way of updating the energy cut-off for planewave discretizations, for both linear and nonlinear eigenvalue problems. The method is {\it error controllable} for linear eigenvalue problems in the sense that for a given required accuracy, an energy cut-off for which the solution matches the target accuracy can be reached efficiently. Further, the method is particularly promising for nonlinear eigenvalue problems in electronic structure calculations as it shall reduce the cost of early iterations in self-consistent algorithms. We present some numerical experiments for both linear and nonlinear eigenvalue problems. In particular, we provide electronic structure calculations for some insulator and metallic systems simulated with Kohn--Sham density functional theory (DFT) and the projector augmented wave (PAW) method, illustrating the efficiency and potential of the algorithm.
\end{abstract}

\begin{keywords}
  Electronic structure calculation, Planewave method, Energy cut-off,
  {\it A posteriori} error estimate, Adaptive algorithm, Projector augmented wave method
\end{keywords}

\begin{AMS}
  65N25, 65N35, 65N15, 35Q40
\end{AMS}

\section{Introduction}
\label{sec:introduction}

The electronic structure modelling of many-particle systems enables the investigation and prediction of properties of molecules and material systems and is used in different fields, such as in chemistry, materials science, biology, and nanosciences.
Most of electronic structure models exhibit (possibly nonlinear) eigenvalue problems, in particular to compute the electronic ground state of a system within the Born--Oppenheimer approximation.
To solve the underlying eigenvalue problems numerically,
the planewave method is one of the most widely used discretization methods~\cite{Kresse1,Payne,Saad}, and is employed in many packages, e.g. ABINIT~\cite{ABINIT}, CASTEP~\cite{CASTEP}, Quantum Espresso~\cite{Espresso}, and VASP~\cite{VASP}. 
By expressing the eigenfunction as a linear combination of planewave functions, this method has the following key advantages~\cite{Saad}.
First, it is a natural discretization for periodic systems in materials science as the planewaves can match the translational symmetry of the systems perfectly.
Second, the transformation from real to reciprocal space and vice versa can be done efficiently with fast Fourier transforms (FFTs), which makes the calculation of the elements of the Hamiltonian, the matrix that needs to be diagonalised, relatively simple. 
Third, as a spectral discretization method, the planewave method  allows to achieve a spectral convergence rate when combined with pseudo-potential approximations~\cite{Kresse2}. 

A key issue in the planewave method is the choice of the energy cut-off, the parameter determining the number of basis functions. 
A larger cut-off will lead to more accurate results, but at the price of increasing the computational cost significantly.
Therefore, it is important to choose an appropriate energy cut-off leading to the right accuracy while using as few planewaves as possible. 
Unfortunately, such a cut-off is {\it a priori} unknown. In practice, it is usually determined by testing the convergence of quantities of interest, such as the energy, the electron density, or the forces and stresses.
Starting any simulation of a new system by a convergence test in order to determine a cut-off satisfying the above criterion can be quite demanding, as this requires to solve several times the same problem for different cut-offs.

The purpose of this paper is to present an elegant algorithm to determine an energy cut-off in an {\it a posteriori} way, so that the planewave approximation reaches the desired accuracy, and avoids superfluous convergence-test computations.
For Schr\"{o}dinger-type linear eigenvalue problems, our algorithm relies on three important aspects:
(i) an {\it a priori} error estimate that gives the decay rate of the discretization error;
(ii) an {\it a posteriori} error estimate that gives guaranteed and asymptotically accurate error indicators;
(iii) the strategy which predicts sensible energy cut-offs during the simulation.
We further incorporate this algorithm into the self-consistent field (SCF) algorithm used to solve nonlinear eigenvalue problems, in particular the Kohn--Sham equations \cite{kohn65,martin05} in Density Functional Theory (DFT) \cite{hohenberg64,kohn65}, so that the energy cut-off is adapted along the iterations of the algorithm.

The idea of adapting the discretization on-the-fly has long been exploited, in particular in finite element methods,
which have been extensively studied for both general elliptic equations \cite{Babuska1, Becker, Binev, Cascon, fler, Garau} and eigenvalue problems in quantum physics \cite{Bao, Chen2, Chen1, Dai6, Motamarri13, Suryanarayana, Tsuchida1}.
In adaptive methods, an efficient and reliable {\it a posteriori} error estimator is essential, in order to indicate how to improve the basis set, e.g. by  modifying the energy cut-off for planewave methods.
To this end, this work will use some guaranteed {\it a posteriori} error estimates presented in~\cite{Cances1,Cances2, Cances:hal-02127954}.
For planewave methods, there are only a handful of works devoted to the construction of the {\it a posteriori} error estimates and adaptive methods. 
We refer to~\cite{Dusson} for {\it a posteriori} error estimates and \cite{Canuto,Gygi} for adaptive algorithms.
To our best knowledge, 
there is very few efficient implementations of adaptive planewave algorithm for quantum eigenvalue problems \cite{Dusson},  and no existing work for simulations of real systems with Kohn--Sham equations.
The difference lies in that the planewaves are non-local basis functions, and the traditional local refinement techniques used in adaptive finite element discretizations cannot be applied straightforwardly. 

As a major application of our algorithm, we carry out electronic structure calculations in the framework of Kohn--Sham DFT and the projector augmented wave (PAW) method \cite{blochl1994projector,fang2019impl,Kresse2}.
The PAW method is a very accurate and widely used method in electronic structure calculations,
which has an elegant theoretical framework based on a linear transformation from the pseudo wave-functions to the real all-electron ones and then derives closed-form expressions for electronic densities, energy functionals, and forces in a consistent manner.  
This method has now been efficiently implemented in several simulation packages, including the well-known VASP and ABINIT codes, and is widely applied in materials science, condensed matter physics, and quantum chemistry. 
In practical PAW calculations, a proper setting of the planewave cut-off for the pseudo wave-function to ensure numerical convergence is critical for obtaining meaningful simulation results.

The rest of the article is organized as follows. 
In \cref{sec:eigen}, we focus on linear Schr\"{o}dinger-type equations. We briefly discuss the planewave discretization, and available {\it a priori} and {\it a posteriori} error estimates. 
In \cref{sec:adaptive:linear}, we propose two strategies to construct an adaptive algorithm for linear eigenvalue problems. 
In \cref{sec:nonlinear}, the Kohn--Sham equations and the PAW method are presented, and the adaptive algorithm for linear eigenvalue problems is built into SCF iterations for the nonlinear eigenvalue problem.
In \cref{sec:numerics}, we perform some numerical experiments to show the validity and efficiency of our algorithms for linear and nonlinear eigenvalue problems.
In particular, Kohn--Sham DFT calculations with the PAW method for some bulk systems are presented, including a diamond system and a metallic FCC aluminum system with a vanishing band gap.
We also provide a detailed comparison with the convergence test approach
(see \cref{sec:numerics_KS_PAW}) to illustrate the efficiency and advantages of the adaptive algorithm.
Finally, some concluding remarks are given in \cref{sec:conclusion}.

Throughout this paper, we shall use $C$ to denote a generic positive constant which may stand for different values at its different occurrences but is independent of finite dimensional subspaces.
The symbol $\langle\cdot,\cdot\rangle$ denotes an abstract duality pairing between a Banach space and its dual, and $(\cdot,\cdot)$ denotes the inner product in the space $L^2_\#(\Omega)$ introduced in \cref{sec:eigen}.

\section{Linear eigenvalue problems}
\label{sec:eigen}

In this section, we consider a Schr\"{o}dinger-type equation, review the planewave discretization and discuss available {\it a priori} and {\it a posteriori} error estimates.
We clarify at this point that, only the error estimates for linear eigenvalue problems are required and exploited for the adaptive planewave algorithms developed in this paper, for both Schr\"{o}dinger-type and Kohn--Sham equations.

Let $d\in\{1,2,3\}$ be the dimension of the system, $\LL = P\Z^d \subset \R^{d}$ be a Bravais lattice, where $P\in \R^{d\times d}$ is a non-singular matrix. 
Let $\Omega$ be a unit cell of the lattice $\LL$, and $\LL'$ be the dual lattice of $\LL$. 
We denote by $e_{\G}({\rr})=|\Omega|^{-1/2} e^{i\G\cdot\rr}$ the planewave with wavevector $\G\in\LL'$. 
The family $\{e_{\G}\}_{\G\in\LL'}$ forms an orthonormal basis set of $L_{\#}^2(\Omega)$, where
\begin{equation*}
	L_{\#}^2(\Omega)=\{u\in L^2_{\rm loc}(\mathbb{R}^{d})~:~u~{\rm is}~\LL{\rm -periodic}\}.
\end{equation*} 

For all $u\in L_{\#}^2(\Omega)$, we have
\begin{equation*}
	u({\rr})=\sum_{\G\in\LL'}\hat{u}_{\G} e_{\G}({\bf r})\quad{\rm with}\quad \hat{u}_{\G}
	= (e_{\G},u)
	=|\Omega|^{-1/2}\int_{\Omega} u({\bf r})e^{-i\G\cdot \rr}\dd{\rr}.
\end{equation*}
We introduce the Sobolev spaces of real-valued $\LL$-periodic functions for $s\in\mathbb{R}$ 
\begin{equation*}
	H_{\#}^s(\Omega)=\left\{u(\rr)=\sum_{\G\in\LL'}\hat{u}_{\G} e_{\G}({\rr}): \forall~{\G}\in\LL', \hat{u}_{-\G} = \hat{u}^{*}_{\G},
	\sum_{{\G}\in\LL'}\big(1+|\G|^2\big)^s |\hat{u}_{\G}|^2<\infty
	\right\} .
\end{equation*}

\subsection{Schr\"{o}dinger-type equations}
\label{sec:schrodinger}

We consider a Schr\"{o}dinger-type linear eigenvalue problem:
Find $(\lambda_i,\varphi_i)\in\R\times H^1_{\#}(\Omega)$, such that $\|\varphi_i\|_{L_{\#}^2(\Omega)}=1$ and
\begin{equation}
    \label{eigen}
    \left(-\Delta+V(\rr)\right) \varphi_i(\rr) =\lambda_i \varphi_i(\rr) 
    \qquad i = 1,2,\cdots,
\end{equation}
where the potential $V\in C^{\infty}(\R^d)$ and is $\LL$-periodic.
The smoothness assumption on the potential $V$ is reasonable, since the planewave discretization is in practice often combined with smooth pseudo-potentials. It implies in particular that $V$ is bounded below.

The weak form of \cref{eigen} reads
\begin{equation}
    \label{eigen:weak}
    a(\varphi_i,v)=\lambda_i(\varphi_i,v) \quad \forall~v\in H^{1}_{\#}(\Omega) \qquad i = 1,2,\cdots
\end{equation}
with the bilinear form $a(\cdot,\cdot):  H^{1}_{\#}(\Omega)\times H^{1}_{\#}(\Omega) \rightarrow \mathbb{R}$ given by
\begin{flalign}
\label{coercevity}
a(u,v)=\int_{\Omega}\nabla u\cdot\nabla v+\int_{\Omega} Vuv .
\end{flalign}
Since the operator $A:= - \Delta +V$ 
is self-adjoint, bounded below with compact resolvent, there exists a non-decreasing sequence of eigenvalues
$\lambda_{1}<\lambda_{2} \leq \cdots \leq \lambda_j \rightarrow +\infty$,
where the $\lambda_j$'s are repeated according to their multiplicity.

Up to shifting the operator by a positive constant, e.g. $\displaystyle \max\big\{1-\inf_{\pmb{r}\in\Omega}V(\pmb{r}),0\big\}$, we can further assume that $V(\pmb{r})\geq 1~(\forall~\pmb{r}\in\Omega)$, which ensures that all eigenvalues of $A$ are positive. Note that such a shifting has no impact on the eigenfunctions of the problem.
Then it holds
\begin{equation*}
a(v,v) \geq \|v\|_{H^1_{\#}(\Omega)}^2 \qquad\forall~v\in H^1_{\#}(\Omega) ,
\end{equation*}
and we can define the corresponding energy norm by $\norma{\cdot}:=\sqrt{a(\cdot,\cdot)}$. 

\subsection{Planewave discretization}
\label{sec:pw}

Given an energy cut-off $\Ec>0$, we define the following finite dimensional space 
\begin{equation}
    \label{def:spaceE}
    X_{\Ec}(\Omega) := \left\{\sum_{\G\in\LL',~|\G|^2\leq 2\Ec} c_{\G}e_{\G}({\rr}), ~ \forall~ \G,~ c_{-\G}=c^{*}_{\G} \right\}.
\end{equation}
For $v\in H^s_{\#}(\Omega)$, the best approximation of $v$ in $X_{\Ec}(\Omega)$ is
\[ \displaystyle \Pi_{\Ec} v=\sum_{\G\in\LL',~|\G|^2\leq 2\Ec} \hat{v}_{\G}
e_{\G}({\bf r})\]
for any $H^t_{\#}$-norm ($t\leq s$).
The more regularity $v$ has, the faster this truncated series converge to~$v$: 
For $s,t\in\R_+$ satisfying $t\leq s$, we have that for each $v\in H^s_{\#}(\Omega)$ (see e.g. \cite{Claudio}),
\begin{equation}
\label{error:interpolation:planewave}
\|v-\Pi_{\Ec} v\|_{H^t_{\#}(\Omega)}=\min_{w\in X_{\Ec}}\|v-w\|_{H^t_{\#}(\Omega)}  \lesssim \Ec^{(t-s)/2}\|v\|_{H^s_{\#}(\Omega)} .
\end{equation}

\medskip

For simplicity of notation, we suppress for now the index for the considered eigenpair in the eigenvalue problem. 
We rewrite the linear problem \cref{eigen:weak} as
\begin{equation}
    \label{eigen:weak_}
    a(\varphi,v)=\lambda(\varphi,v) \qquad \forall~ v\in H^{1}_{\#}(\Omega),
\end{equation}
and present available {\it a priori} and {\it a posteriori} error estimates.
We note that the following discussions are not restricted to a specific eigenpair, but require a gap between the considered eigenvalue and the surrounding ones.

Given an energy cut-off $\Ec$, the planewave discretization is a Galerkin approximation of the eigenpairs of \cref{eigen:weak_} within the finite dimensional subspace $X_{\Ec}(\Omega)\subset H_{\#}^1(\Omega)$:
Find $(\lambda_{\Ec}, \varphi_{\Ec})\in\R\times X_{\Ec}(\Omega)$, such that $\|\varphi_{\Ec}\|_{L^2_\#(\Omega)}=1$ and
\begin{equation}
    \label{eigen:planewave}
    a(\varphi_{\Ec},v) = \lambda_{\Ec}(\varphi_{\Ec},v) \qquad \forall~ v\in X_{\Ec}(\Omega) .
\end{equation}

\subsection{{\it A priori} error estimate}
\label{sec:priori}

An {\it a priori} estimate for the solutions $(\lambda_{\Ec}, \varphi_{\Ec})$ of \cref{eigen:planewave} can be found in \cite[Theorems 2 and 3]{osborn75}.
There exists a constant $C>0$ independent of $\Ec$, such that
\begin{align}
\label{err:priori:eigenfunction}
& \|\varphi-\varphi_{\Ec}\|_{H^{1}_{\#}(\Omega)} \leq C \inf_{\psi_{\Ec}\in X_{\Ec}} \|\varphi- \psi_{\Ec}\|_{H^{1}_{\#}(\Omega)} 
\qquad{\rm and}
\\[1ex]
\label{err:priori:eigenvalue}
& |\lambda-\lambda_{\Ec}| \leq C\|\varphi-\varphi_{\Ec}\|_{H^{1}_{\#}(\Omega)}^2 .
\end{align}
Hence, if the eigenfunction $\varphi$ is analytic,
which we will assume in the following,
then the approximation error decays exponentially as (see \cite[Section 5.4, Eq. (5.4.5)]{Claudio})
\begin{align}
\label{err:exponential:phi}
& \|\varphi-\varphi_{\Ec}\|_{H^{1}_{\#}(\Omega)} \leq Ce^{-c\sqrt{\Ec}}
\qquad{\rm and}
\\[1ex]
\label{err:exponential:lambda}
& \big|\lambda-\lambda_{\Ec}\big| \leq Ce^{-\tilde{c}\sqrt{\Ec}},
\end{align}
with the constants $C, c, \tilde{c}>0$ independent of $\Ec$. 
The above {\it a priori} error estimates not only guarantee that the error goes to zero as the energy cut-off increases, but also offer the speed of convergence of the approximate solutions to the exact one.

We observe from the numerical experiments (see \cref{fig:linear:err} in \cref{sec:testPrinciples}) that for the smooth tested potentials,
the planewave approximation errors decay exponentially fast with respect to $\sqrt{\Ec}$, which matches the {\it a priori} error estimates in \cref{err:exponential:phi,err:exponential:lambda}.

In \cref{sec:adaptive:linear}, we will combine the {\it a priori} bounds \cref{err:exponential:phi,err:exponential:lambda}
with the {\it a posteriori} bounds presented below in a linear regression strategy for determining good energy cut-offs.

\begin{remark}
	Analogous to the linear eigenvalue problem \cref{eigen:weak_}, some optimal {\it a\ priori} error estimates for nonlinear eigenvalue problems can be obtained \cite{Cances3,cances12,chen13}. 
\end{remark}

\subsection{{\it A posteriori} error estimate}
\label{sec:posteriori}

To develop an adaptive method, it is very useful to have {\it a posteriori} error estimators available.
Unlike {\it a priori} error estimates, the goal of the {\it a posteriori} error estimates is to provide a computable bound on the error between the numerical approximation of the solution and the unknown exact solution. 
This bound should therefore be computable with the knowledge of the approximate solution and the parameters used for the computation only.

In this section, we review an {\it a posteriori} error estimator for the planewave approximation of the linear eigenvalue problem \cref{eigen},
which gives computable and asymptotically accurate approximate errors.
This {\it a posteriori} error indicator was first introduced for simple eigenvalues for conforming finite element discretizations in \cite{Cances1}, nonconforming methods in \cite{Cances2}, and then extended to eigenvalue clusters in \cite{Cances:hal-02127954}. 

Let us denote $H^{1}_{\#}(\Omega)$ by $H$ and its dual by  $H'$ for simplicity of notations.
Let $(\lambda,\varphi)\in\R\times H$ be an eigenpair of \cref{eigen:weak_} and $(\lambda_{\Ec},\varphi_{\Ec})\in\R\times X_{\Ec}(\Omega)$ be the corresponding planewave approximation with a given energy cut-off $\Ec$.
We define the residual $\res(\lambda_{\Ec},\varphi_{\Ec}) \in H'$ by
\begin{equation}
    \label{residual}
    \big\langle\res(\lambda_{\Ec},\varphi_{\Ec}) ,v\big\rangle_{H',H} = a\big(\varphi_{\Ec},v\big)-\lambda_{\Ec}\big(\varphi_{\Ec},v\big) \qquad \forall~v\in H.
\end{equation}
The idea of residual-based {\it a posteriori} error estimation is to estimate the approximation error
by the dual norm of the residual in an appropriate Hilbert space. 
Note that the dual norm must be well chosen, otherwise the norm of the residual might not accurately represent the error or not even go to zero when the error goes to zero. 
To construct the {\it a posteriori} error estimator for the approximation $(\lambda_{\Ec},\varphi_{\Ec})$, we consider the dual norm of the residual
\begin{equation}
    \label{error:indicator}
    \eta(\lambda_{\Ec},\varphi_{\Ec}) := \normd{\res(\lambda_{\Ec},\varphi_{\Ec})},
\end{equation}
where the dual norm (with respect to the energy norm $\norma{\cdot}$) is defined for $f\in H'$ by
\begin{flalign*}
\normd{f}:= \sup_{v\in H,~v\neq 0} \frac{\langle f,v\rangle_{H',H}}{\norma{v}}. 
\end{flalign*}
For simplicity of notation, we will denote the residual and its dual norm by
\begin{equation*}
	\res_{\Ec} = \res(\lambda_{\Ec},\varphi_{\Ec})
	\qquad{\rm and}\qquad
	\eta_{\Ec} = \eta(\lambda_{\Ec},\varphi_{\Ec}) 
\end{equation*}
in the following.
We have from \cite{Cances:hal-02127954} that if $\lambda$ is a simple eigenvalue, there exists $\alpha(\Ec)$ with $\alpha(\Ec) \rightarrow 0$ as $\Ec \rightarrow \infty$ such that
\begin{flalign}
\label{error:posteriori:estimate}
0 \le 	\lambda_{\Ec}-\lambda \leq \big(1 + \alpha(\Ec)\big)\eta_{\Ec}^2.
\end{flalign}

\begin{remark}\label{remark:multiple}
	So far, we have ignored the index $j$ for the eigenpairs for simplicity of notation, but the error indicator for a specific eigenpair with index $j$ can also be denoted by $\eta_{\Ec,j}$ for completeness.
	Thus, when considering multiple eigenpairs, the definition of $\eta_{\Ec}$ can be extended by taking the sum
	\begin{equation*}
		\eta_{\Ec} := \left( \sum_{j\in J} \eta_{\Ec,j}^2 \right)^{1/2} ,
	\end{equation*}
	where $J$ is the set of indices of eigenpairs under consideration.
	Moreover, if the quantity of interest is not the sum of the eigenvalues, but a weighted sum, which happens to be the case in practice when fractional occupation numbers are in use, 
	one can obtain error indicators on this quantity of interest by introducing corresponding weights in front of the individual error indicators $\eta_{\Ec,j}^2$, as defined in \cref{sec:adaptive:scf}.
\end{remark}

\begin{remark}\label{remark:nonlocal}
	The above {\it a posteriori} error estimate can be directly generalized to linear operators with non-local potentials (see e.g.  \cref{v:nonlocal}).
\end{remark}

To evaluate the {\it a posteriori} error indicator \cref{error:indicator}, one needs to solve the infinite dimensional linear system \cref{residual} to compute the residual.
A practical evaluation method is to restrict the trial function space and test function space in \cref{residual} to a finite dimensional subspace $X_{\Eg}(\Omega)$ (see definition \cref{def:spaceE})
with some cut-off $\Eg\gg\Ec$.
We will then denote the computable error indicator by $\eta_{\Ec}^{\Eg}$, whose detailed calculations are presented in \cref{sec:evaluation}.
If $\Eg$ is proportional to $\Ec$, then a standard calculation by solving the corresponding finite dimensional linear system will require a cost of $\mathcal{O}(\Ec^{d/2}\ln\Ec)$, and we refer to \cref{sec:standard_eta} for the explanations.

In order to reduce the cost of the error indicator, we can alternatively evaluate the residual via a perturbation-based method. The resulting error indicator will be denoted by $\eta_{\Ec}^{\Eg,[k]}$ (with $k=1,2$ the order of perturbation), whose cost is reduced by avoiding 
the resolution of a linear system.
We refer to \cref{sec:perturbation_eta} for the derivation of this method and discussion of its computational cost.

\section{Adaptive algorithm for linear problems}
\label{sec:adaptive:linear}

We are now ready to construct the adaptive algorithm for Schr\"{o}dinger-type equations. 
The algorithm repeats the following procedure until the required accuracy is reached: 
\begin{itemize}
	\item[(a)]
	given the energy cut-off $\Ec$, solve \cref{eigen:planewave} to obtain the approximate eigenpair $(\lambda_{\Ec},\varphi_{\Ec})$;
	\item[(b)] 
	compute the {\it a posteriori} error estimator
	$\eta_{\Ec}^{\Eg}$ (or $\eta_{\Ec}^{\Eg,[k]}$)
	from the approximation $(\lambda_{\Ec},\varphi_{\Ec})$ and a given $\Eg$ (or $\Eg$ and $k$);
	\item[(c)] 
	use some strategy to determine the new energy cut-off $\Ecopt$.
\end{itemize}

The step (a) is standard and the step (b) has been discussed in the previous section.
To determine the energy cut-offs during the iterations, i.e. step (c),  
we propose two strategies.
Note that to initialize the algorithm, we choose a relatively small energy cut-off at the first step.

\begin{figure}[htb!]
        \centering 
		{\includegraphics[width=0.45\textwidth]{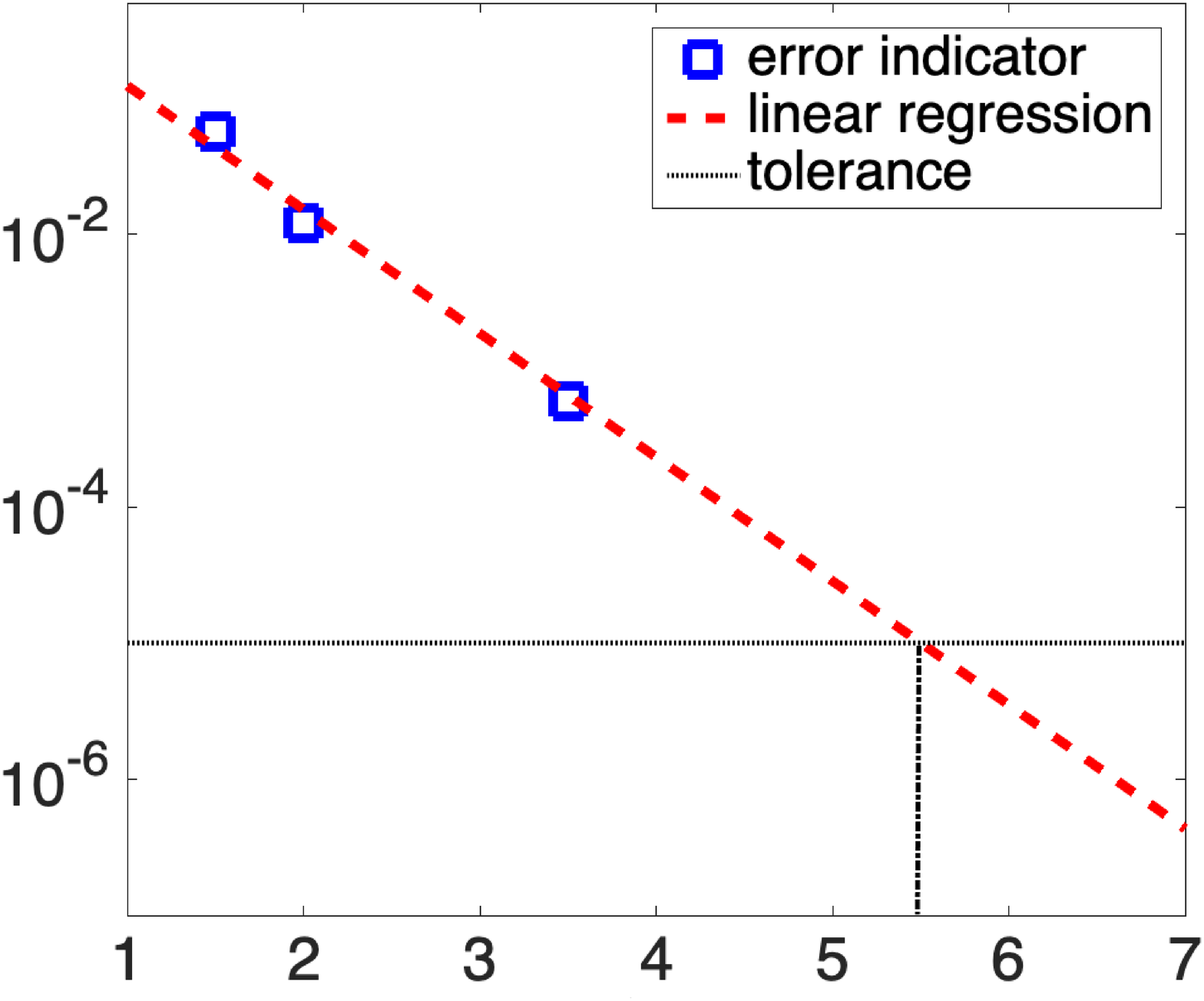}}
        \put(-60,36){\small \pmb{$\sqrt{\EcA^{(l+1)}}$}}
        \hskip 0.2cm
		{\includegraphics[width=0.37\textwidth]{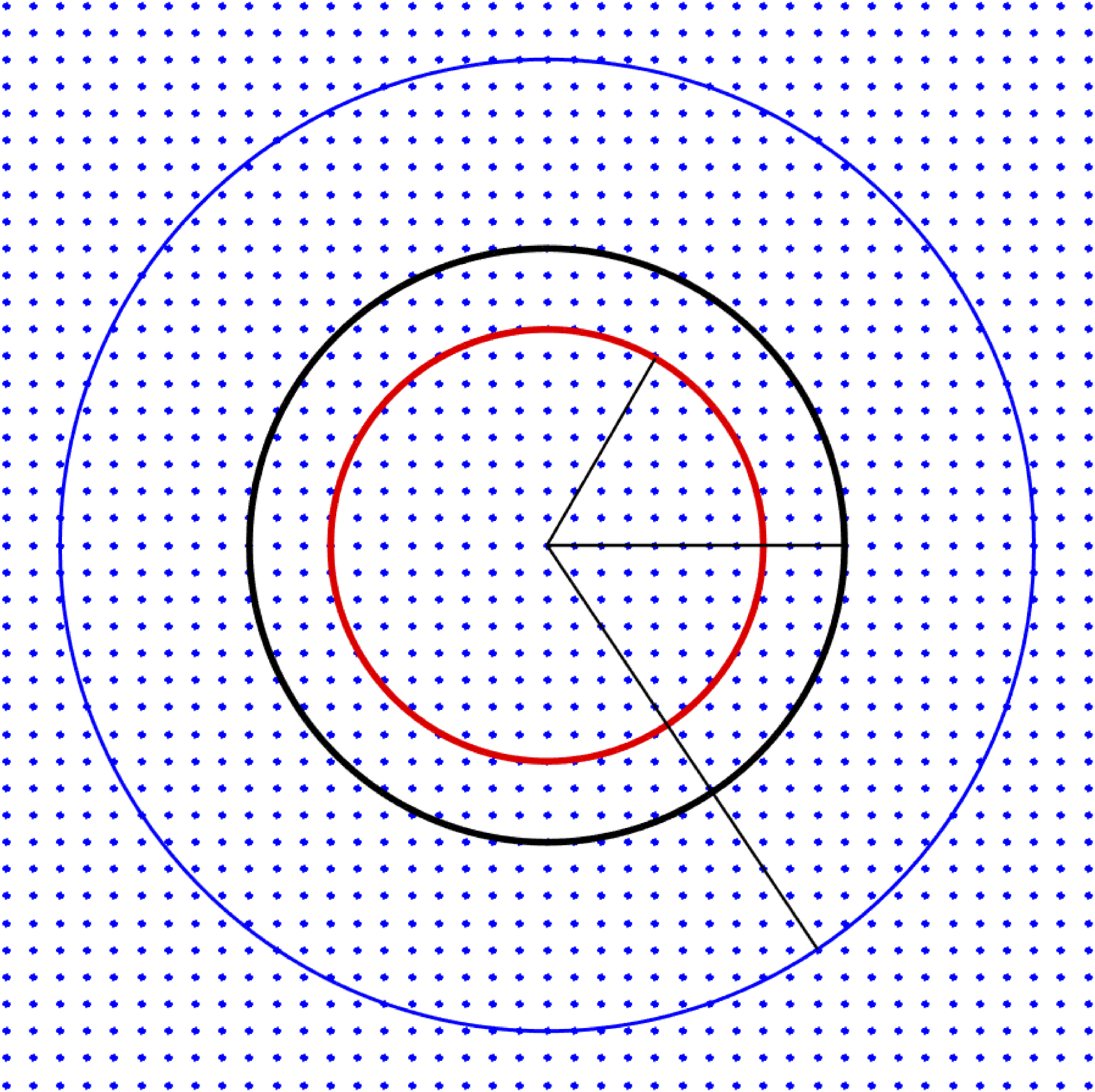}}
		\put(-115,115){\pmb{$\mathbb{L}'$}}
		\put(-85,80){\small \pmb{$\sqrt{2\Ec^{(l)}}$}}
		\put(-50,70){\small \pmb{$\sqrt{2\EcB^{(l+1)}}$}}
		\put(-65,35){\small \pmb{$\sqrt{2\Eg}$}}
    \caption{Strategy A (left) and B (right) for determining the energy cut-offs.}
	\label{fig:strategy}
\end{figure}

The first strategy relying on the known {\it a priori} spectral convergence rate (in \cref{sec:priori}) and the {\it a posteriori} error estimate (in \cref{sec:posteriori}) is presented in Strategy A.
The idea is to perform a linear regression to predict the energy cut-off leading to a required accuracy. As an input for the linear regression, we use the $l$-th energy cut-offs and the {\it a posteriori} error indicators $\big\{\Ec^{(i)},\eta_{\Ec^{(i)}}^{\Eg} ({\rm or} \ \eta_{\Ec^{(i)}}^{\Eg,[k]}) \big\}_{0\leq i\leq l}$.
We show in \cref{fig:strategy} a schematic plot of Strategy A.
\begin{algorithm}[H]
	\setstretch{0.5}
	\caption*{{\bf Strategy A:} Linear regression}
	\begin{algorithmic}
		\vskip 0.1cm
		\REQUIRE
		energy tolerance $\varepsilon$, 
		energy cut-offs and error estimators $\big\{\Ec^{(i)},\eta_{\Ec^{(i)}}^{\Eg}\big\}_{0\leq i\leq l}$.
		\begin{enumerate}
			\item 
			Perform a linear regression ($ {\textbf L \textbf R}$) to compute the linear function
			\begin{equation}
			\label{func:LR}
			\Theta(x) := {\textbf L \textbf R} \left[ \left\{ \sqrt{\Ec^{(i)} }, \log\big(\eta_{\Ec^{(i)}}^{\Eg}\big) \right\} _{0\leq i\leq l}  \right] (x) .
			\end{equation}
			\item  
			Solve the following linear equation to obtain $\bar{E}$
			\begin{equation*}
				\Theta(\bar{E}) = \frac{1}{2}\log\varepsilon.
			\end{equation*}
		\end{enumerate}
		\ENSURE 
		Output energy cut-off $\EcA^{(l+1)}	= \bar{E}$
	\end{algorithmic}  
\end{algorithm}

In \cref{func:LR}, {\textbf L\textbf R} gives a linear function $\Theta(x)=ax+b$, where the coefficients $a$ and $b$ are determined by least squares  as follows
\begin{equation}
\label{LR:least_square}
\min_{a,b} \left\{ \sum_{0\leq i\leq l} \Big| a\sqrt{\Ec^{(i)}} + b - \log\big(\eta_{\Ec^{(i)}}^{\Eg}\big) \Big|^2 \right\}.
\end{equation}

We observe from the numerical results in \cref{sec:numerics} that 
the strategy based on linear regression appears to be sharp in many cases, but sometimes overestimates the optimal energy cut-offs due to the fluctuation of the errors.

A second strategy is presented in Strategy B below.
This strategy is analog to the D\"{o}rlfer strategy, which is a widely used method in adaptive finite elements methods.
In this method, we define an error indicator $\eta^{E_g}_{E_c^{(l)}}(E)$ composed of all contributions in $\eta^{E_g}_{E_c^{(l)}}$ from all wave vectors $\G$ in the reciprocal space $\Ec\leq\frac{1}{2}|{\G}|^2\leq E$, i.e.
\[
    \eta^{E_g}_{E_c^{(l)}}(E) =  
    \left(
\sum_{{\G}\in \LL',~\Ec\leq\frac{1}{2}|{\G}|^2\leq E}
\left| \widehat{\eta^{E_g}_{E_c^{(l)}, \G}} \right|^2
\right)^{1/2}.
\]
More details are given in Appendix \ref{sec:perturbation_eta}, more precisely, in \cref{err:res:local:E} and \cref{err:local:perturb}.
The energy cut-off $E$ is then increased until sufficiently many $\G$'s are included so that the chosen tolerance is achieved.

\begin{algorithm}[H]
	\setstretch{0.5}
	\caption*{{\bf Strategy B:} Error reduction}
	\begin{algorithmic}
		\vskip 0.1cm
		\REQUIRE 
		eigenvalue tolerance  $\varepsilon>0$, 
		energy cut-off $\Ec^{(l)}$
		and error estimator $\eta_{\Ec^{(l)}}^{\Eg}$.
		\begin{enumerate}
			\item[]
			Find $\bar{E}>\Ec^{(l)}$ such that $\bar{E}$ is the minimal value satisfying
			\begin{equation}
			\label{Ec:strategy1}
			\eta_{\Ec^{(l)}}^{\Eg}(\bar{E}) \geq  \sqrt{\left(\eta_{\Ec^{(l)}}^{\Eg}\right)^2 - \varepsilon} .
			\end{equation}
		\end{enumerate}
		\ENSURE 
		Output energy cut-off $\EcB^{(l+1)} = \bar{E}$
	\end{algorithmic}  
\end{algorithm}

We show in \cref{fig:strategy} a schematic plot of Strategy B. 
In the picture, the red and blue circles represent the cut-offs $\Ec$ and $\Eg$ respectively, and the {\it a posteriori} error estimate consists of the sum over all reciprocal lattice vectors that lie between these two circles. Strategy B determines the black circle, which is obtained so that the planewave vectors lying between the red and black ones contribute to ``most of the error" up to some given tolerance.
In practice, \cref{Ec:strategy1} can be solved by using the bi-section or golden-section method to obtain $\bar{E}$.
We refer to \cite{Canuto,Pan18} for discussions related to the D\"{o}rlfer strategy for planewave approximations, in which no practical algorithm was implemented.

From a practical point of view, it is more stable and reliable to use a combination of the above two strategies, which we describe in Algorithm 1 below. In terms of strategy, taking the minimum  between the cutoffs obtained from Strategy A and B (as in~\cref{Ec:algorith1} below) is expected to minimize the risk of overestimating the new energy cut-off. Alternatively, one could use `max' in \cref{Ec:algorith1} instead of `min' to minimize the number of iteration steps, and hence the number of calls to the linear eigensolver in the algorithm.

\begin{algorithm}[H]
	\setstretch{0.6}
	\caption*{{\bf Algorithm 1:} Adaptive planewave method for linear eigenvalue problems}
	\begin{algorithmic}
		\vskip 0.1cm
		\REQUIRE 
		eigenvalue tolerance $\varepsilon>0$, 
		initial energy cut-off $\Ec^{(0)}$,
		$\Eg\gg 1$	and $l=0$.
		\begin{enumerate}
			\item 
			Solve \cref{eigen:planewave} within $X_{\Ec^{(l)}}$ to obtain the planewave approximation $\big(\lambda_{\Ec^{(l)}} , \varphi_{\Ec^{(l)}} \big)$.
			\item 
			Compute the {\it a posteriori} error estimator defined by \cref{err:computable:Eg} or \cref{err:computable:Eg-k}.
			\item 
			If $\left(\eta_{\Ec^{(l)}}^{\Eg} \right)^2 < \varepsilon$, then stop and return the planewave approximation. 
			Else, goto 4.
			\item 
			If $l=0$, then use Strategy B to get $\Ec^{(l+1)}=\EcB^{(l+1)}$, let $l=l+1$ and goto 2.
			
			Else, use Strategy A to compute $\EcA^{(l+1)}$ and Strategy B to compute $\EcB^{(l+1)}$. 
			Take
			\begin{equation}
			\label{Ec:algorith1}
			\Ec^{(l+1)} = 	\min\left\{ \EcA^{(l+1)} , \EcB^{(l+1)} \right\}
			\end{equation}
			and $l=l+1$, goto 1.
		\end{enumerate}
		\ENSURE 
		$\left( \lambda_{\Ec^{(l)}} , \varphi_{\Ec^{(l)}}  \right)$
	\end{algorithmic}  
\end{algorithm}

Note that the computational cost for both error indicators and Strategy B increase with respect to the cut-off $\Eg$, but we can adjust $\Eg$ during the adaptive algorithm according to the energy cut-off $\Ec$ to avoid using very large $\Eg$ throughout the algorithm. 

Let us now comment on the computational cost of the algorithm, assuming that we need to compute $N_{\rm b}$ eigenvalues. 
At each step of Algorithm 1, the cost for solving the eigenvalue problem is $\mathcal{O}(N_{\rm b}^2\Ec^{d/2})$, the cost for evaluating the error indicators is $\mathcal{O}(N_{\rm b}\Eg^{d/2}\ln\Eg)$.
Therefore, if $\Eg$ stays relatively small and comparable to $\Ec$ then the overall computational costs of Algorithm 1 is $\mathcal{O}(N_{\rm b}^2\bar{E}_{\rm c}^{d/2})$, where $\bar{\Ec}$ is the optimal energy cut-off determined by the algorithm.
Hence, the cost of the whole procedure is still dominated by the linear eigensolver. 
Compared to a classical convergence test, this procedure 
does not require user's experience to find a good energy cut-off.

\section{Applications to nonlinear eigenvalue problems}
\label{sec:nonlinear}

In this section, we will introduce an adaptive planewave method for nonlinear eigenvalue problems.
We will first briefly review the Kohn--Sham equations, a widely used model in electronic structure calculations as well as its formulation in the framework of the projector augmented wave (PAW) method.
We will then discuss how the adaptive algorithms developed for linear eigenvalue problems can be generalized to these nonlinear eigenvalue problems.
We mention that only the error estimates for linear eigenvalue problems are required for the algorithms presented in this section.

\subsection{Kohn--Sham equations}
\label{sec:ks}

In the {\it ab-initio} quantum mechanical modeling of many electron systems, 
the Kohn--Sham density functional theory (DFT) \cite{hohenberg64,kohn65} has established itself as one of the most
powerful electronic structure methods due to its good balance between accuracy and computational cost.

We consider a many-particle system (with $\Nn$ nuclei and $\Ne$ electrons) in the spin-less and non-relativistic setting. 
The Kohn--Sham ground state solution can be obtained by solving the following Kohn--Sham equations \cite{martin05}:
Find $(\lambda_i,\varphi_i)\in\R\times H_{\#}^1(\Omega),~i=1,\cdots,\Nb$, such that $ \int_{\Omega}\varphi_i\varphi_j = \delta_{ij}$ and
\begin{equation}
\label{eq:ks-standard}
H[\rho]\varphi_i = \lambda_i\varphi_i ,
\end{equation}
where $\Nb$ is the number of eigenpairs/bands under consideration, $\{\lambda_i\}_{i=1}^{\Nb}$ are the $\Nb$ lowest eigenvalues, 
$\rho=\sum_{i=1}^{\Nb}f_i|\varphi_i|^2$ is the electron density with $f_i$ the occupancy number, and 
\begin{flalign}\label{ks:H}
H[\rho] = -\frac{1}{2}\Delta+\Veff[\rho]
\qquad{\rm and}\qquad
\Veff[\rho] = v_{\rm ps} + \VH[\rho] + \Vxc[\rho]
\end{flalign}
with $\Veff[\rho]$ the effective potential, $\Vps$ the pseudo-potential generated by the nuclei and core electrons,  $\VH[\rho]$ and $\Vxc[\rho]$ the so-called Hartree potential and exchange-correlation potential, respectively.
The occupancy numbers $f_i$ can be chosen by smearing methods like the
Fermi--Dirac or the Methfessel--Paxton scheme.

We shall now give more details on the potentials used to define $V_{\rm eff}$.
In the norm-conserving pseudo-potential setting \cite{martin05}, the pseudo-potential $\Vps$ can be written as
\begin{equation*}
	\Vps = v_{\rm loc} + v_{\rm nl} ,
\end{equation*}
where $v_{\rm loc}:\R^d\rightarrow\R$ is the local part,  $v_{\rm nl}$ is a non-local operator given by
\begin{equation}
    \label{v:nonlocal}
    \big(v_{\rm nl}\phi\big)({\bf r}) = \sum_{I=1}^{\Nn}\sum_{j=1}^{M_{\rm ps}} (\phi,\xi_{I,j})\xi_{I,j}({\bf r}) \qquad\forall~\phi\in L_{\#}^2(\Omega) 
\end{equation}
with $M_{\rm ps}\in\Z_+$ and  $\xi_{I,j}~(I=1,\cdots,\Nn,~j=1,\cdots,M_{\rm ps})\in L_{\#}^2(\Omega)$ being sufficiently smooth functions in $\R^d$.
In the context of
ultra-soft pseudo-potentials (USPP) or PAW methods \cite{laasonen1993car,vanderbilt1990soft}, the non-local parts are formulated by	
\begin{equation}
\label{v:nlocUCPPPAW}
\big(v_{\rm nl}\varphi\big)({\bf r}) = \sum_{I=1}^{\Nn}\sum_{n,m=1}^{M_{\rm ps}} D_{nm}^I (\beta_{I,m},\varphi) \beta_{I,n}({\bf r}) \qquad\forall~\varphi\in L_{\#}^2(\Omega) ,
\end{equation}
where $D_{nm}^I$ depends on the eigenfunctions and need to be updated during the SCF loop. For more details on the PAW method, which  will be the model used in practice for the numerical experiments in \cref{sec:numerics}, see \cref{sec:PAW}.

In periodic systems, $\VH[\rho]$ represents the $\LL$-periodic Coulomb potential generated by the $\LL$-periodic electron density $\rho$
\begin{equation*}
	\VH[\rho]({\bf r}) = 4\pi\sum_{{\G}\in\LL'\backslash\{\bf 0\}} |\G|^{-2} \hat{\rho}_{\G}e_{\G}({\rr}) .
\end{equation*}
Finally, for the exchange-correlation potential, we use a generalized gradient approximation (GGA) \cite{pbe96} in our numerical experiments.

Problem \cref{eq:ks-standard} is a nonlinear eigenvalue problem since the operator $H[\rho]$ depends on the electron density $\rho$, and hence on the eigenfunctions $\{\varphi_i\}_{i=1}^{\Nb}$.
A self-consistent field (SCF) iteration algorithm \cite{martin05} is commonly resorted to for this kind of nonlinear eigenvalue problem. 
At each iteration, a new Hamiltonian is constructed from a trial electronic state and a linear eigenvalue problem,
in the form of Schr\"{o}dinger-type equation~\cref{eigen}, is then solved to obtain the low-lying eigenvalues and corresponding eigenvectors. 

Once the ground state solution of Kohn--Sham equations have be obtained, we can evaluate the band structure energy $E_\uf{bands} \equiv \sum_{i=1}^{N_b} f_i\lambda_{i}$ as a weighted sum of the eigenvalues by occupation numbers. When we consider the $k$-point sampling, the integration over the Brillouin zone also needs to be included.

In the first iterations of the self-consistent field algorithm, the iterates are far form the exact solution, so that the error coming from the iterative solver is dominant. But close to self-consistency, the choice of the basis set becomes important, as the basis set size will ultimately determine the quality of the final approximation.
Therefore, an efficient planewave SCF algorithm should be able to adjust the energy cut-off on the fly, so that no computational resource is wasted while the required accuracy can be reached at the end of the iterations.
We refer to \cite{Dusson} for a similar discussion of error balancing for the Gross-Pitaevskii equation.

\subsection{Adaptive algorithm for nonlinear problems}
\label{sec:adaptive:scf}

We now discuss the construction of error indicators and the adaptive algorithm for the nonlinear Kohn--Sham equations~\cref{eq:ks-standard}.
As mentioned in \cref{sec:ks}, an SCF iteration algorithm is used to solve the nonlinear eigenvalue problem, in which a linear Schr\"{o}dinger-type equation is solved at each step.
Therefore, we can incorporate Algorithm 1 (for linear problems) into each SCF iteration, and thus adapt the energy cut-off on the fly with some well-chosen tolerance, corresponding to the self-consistency error.
The use of the self-consistency error as the tolerance for adapting the cut-off is based on the following: when the solution is far from self-consistency, 
one can choose a relative small energy cut-off for the linearized eigenvalue problem; when the iteration is close to self-consistency,
one has to use a large energy cut-off to obtain accurate eigenpairs.

Let $\rho_{\rm in}^{(m)}$ and  $\rho_{\rm out}^{(m)}$ represent the input and output electron densities at the $m$-th step of the SCF algorithm.
At the $m$-th step, a linearized eigenvalue problem, with the trial input electron density $\rho_{\rm in}^{(m)}$ is solved
\begin{flalign}
\label{eigen:SCF-m}
H\big[\rho_{\rm in}^{(m)}\big] \varphi_i^{(m)}  = \lambda_i^{(m)}  \varphi_i^{(m)}  \qquad i=1,2,\cdots \Nb
\end{flalign}
with some energy cut-off $\Ec^{(m)} $.
Then an output electron density $\rho_{\rm out}^{(m)}$ is obtained from the eigenfunctions $\big\{\varphi_i^{(m)}\big\}_{i=1}^{\Nb}$.
The self-consistency error is defined by
\begin{equation}\label{err:SCF}
\eta_{\rm SCF}^{(m)} := \alpha^{1/2}
\|\rho_{\rm in}^{(m)}-\rho_{\rm out}^{(m)}\|_{L^2(\Omega)} 
\qquad{\rm for~some}~\alpha>0.
\end{equation}
The parameter $\alpha$ in \cref{err:SCF} is set to control the ratio between discretization error and self-consistency error in the total error.
In our numerical simulations, $\alpha$ is set empirically: we take $\alpha$ to be the ratio between the tolerances of the discretization error and self-consistency error. 
Note that a good choice of $\alpha$ could be obtained by a careful error analysis of the nonlinear eigenvalue problems, estimating  explicit optimal weights (in the total error) of the discretization error and self-consistent error.
This will be investigated in future work.

The planewave discretization error indicator is defined with respect to the linear eigenvalue problem \cref{eigen:SCF-m}, which involves all $\Nb$ eigenpairs under consideration, as discussed in \cref{remark:multiple}.
In analogy with definition~\cref{error:indicator}, we define the discretization error indicator as
\begin{equation}
\label{err:indicator:SCF}
\eta_{\Ec}^{(m)} := \left( \sum_{i=1}^{\Nb} f_i \normd{\res^{(m)}_i}^2 \right)^{1/2}
\end{equation}
where $f_i$ is the occupancy number, $\|\cdot\|_{a'}$ is the dual norm corresponding to the linear operator 
\begin{equation*}
	A^{(m)}:=H\big[\rho_{\rm in}^{(m)}\big],
\end{equation*}
and $\res^{(m)}_i$ is the residual for the $i$-th eigenpair at the $m$-th step
\begin{equation}
\label{res:ks}
\res^{(m)}_i := H\big[\rho_{\rm in}^{(m)}\big] \varphi_i^{(m)} - \lambda_i^{(m)} \varphi_i^{(m)}  ~\in~ H' .
\end{equation}
With a given $\Eg\gg\Ec^{(m)}$, we define the computable discretization error indicator by
\begin{equation}
\label{error:SCF-m-standard}
\eta_{\Ec}^{\Eg, (m)} := \sum_{i=1}^{\Nb} f_i \left( \sum_{{\G}\in\LL',~\Ec\leq\frac{1}{2}|{\G}|^2\leq\Eg} \widehat{\res_i^{(m)}}^{\Eg}_{\G} \cdot \widehat{A^{(m),-1}\res_i^{(m)}}^{\Eg}_{\G} \right)^{\frac{1}{2}}
\end{equation}
based on the standard calculation \cref{err:computable:Eg}, and for $k=1,2$.
\begin{align}
\label{error:SCF-m-perturbation}
\eta_{\Ec}^{\Eg, (m),[k]} := \sum_{i=1}^{\Nb} f_i \left( \sum_{{\G}\in\LL',~\Ec\leq\frac{1}{2}|{\G}|^2\leq\Eg} \widehat{\res_i^{(m)}}^{\Eg}_{\G} \cdot \widehat{A^{(m),-1,[k]}\res_i^{(m)}}^{\Eg}_{\G} \right)^{\frac{1}{2}} ,
\end{align}
based on a perturbation method \cref{err:computable:Eg-k}.

Combining the SCF algorithm and Algorithm 1 for linear eigenvalue problems, we propose the following adaptive algorithm for nonlinear eigenvalue problems.

\begin{algorithm}[H]
	\setstretch{0.6}
	\caption*{{\bf Algorithm 2:} Adaptive planewave method for nonlinear eigenvalue problems}
	\begin{algorithmic}
		\vskip 0.1cm
		\REQUIRE 
		tolerance $\delta>0$, 
		initial energy cut-off $\Ec^{(1)}$,
		$\Eg\gg 1$,
		$m=1$ 
		and $\rho^{(0)}\in L^2_{\#}(\Omega)$
		\begin{enumerate}
			\item 
			Solve \cref{eigen:SCF-m} with the trial electron density $\rho^{(m-1)}$ within $X_{\Ec^{(m)}}$ to obtain the planewave approximations $\Big\{ (\lambda^{(m)}_{j,\Ec^{(m)}} , \varphi^{(m)}_{j,\Ec^{(m)}} ) \Big\}_{j=1}^{\Nb}$.
			Compute the error indicator $\eta_{\rm SCF}^{(m)}$.
			\item 
			Compute the error indicator $\eta_{\Ec^{(m)}}^{\Eg,(m)}$ \big(or $\eta_{\Ec^{(m)}}^{\Eg,(m),[k]}$\big).
			\item 
			If $\left(\eta_{\rm SCF}^{(m)}\right)^2 < \delta$ and $\left(\eta_{\Ec^{(m)}}^{\Eg,(m)}\right)^2<\delta$, then stop and return the approximations.
			Else, goto 4.
			\item 
			If $\left(\eta_{\Ec^{(m)}}^{\Eg,(m)}\right)^2< \left(\eta_{\rm SCF}^{(m)}\right)^2$, then goto 5.
			
			Else, use Algorithm 1 to solve \cref{eigen:SCF-m} with tolerance $\varepsilon = \left({\eta_{\rm SCF}^{(m)}}\right)^2$,
			let $\Ec^{(m)}$ be the final energy cut-off of Algorithm 1, goto 5.
			\item
			Let $\displaystyle \rho^{(m)}$ be the electron density from $\Big\{\varphi^{(m)}_{j,\Ec^{(m)}}\Big\}_{i=1}^{\Nb}$ and $m=m+1$, goto 1.
		\end{enumerate}
		\ENSURE Approximate eigenfunctions and eigenvalues
		$\Big\{ (\lambda^{(m)}_{j,\Ec^{(m)}} , \varphi^{(m)}_{j,\Ec^{(m)}} ) \Big\}_{j=1}^{\Nb}$
	\end{algorithmic}  
\end{algorithm}

In Step 4 of Algorithm 2, by taking $\left(\eta_{\rm SCF}^{(m)}\right)^2$ as the tolerance for planewave discretizations, we are actually estimating
the total energy error of the full nonlinear eigenvalue problem by the sum $\left(\eta_{\Ec}^{(m)}\right)^2 + \left(\eta_{\rm SCF}^{(m)}\right)^2$.
We note that it could be possible to construct some {\it a posteriori} error estimate for the nonlinear eigenvalue problems directly (see  \cite{Cances4,Chen2,Chen1,Dusson}).
However, we will not use or discuss this type of estimates in this work.

In Step 5 of Algorithm 2, one can exploit a charge mixing scheme to ensure or accelerate the convergence of the SCF algorithm.
The charge mixing in our implementation is performed in the Fourier space.
Assume that we need to mix two charge densities $\rho_1$ and $\rho_2$ with cut-offs $\Ec^1$ and $\Ec^2$, respectively, and $\Ec^2>\Ec^1$. 
We simply fill the rest of planewave components of $\rho_1$ by zeros to expand it to the same grid as $\rho_2$ and then carry out the mixing. 
The additional computational cost of this implementation is negligible.

The most important feature of Algorithm 2 is that it is adaptive.
It automatically determines a good discretization parameter, in our case the energy cut-off, on the fly to achieve the required accuracy. 
Moreover, compared to the conventional SCF iterations, Algorithm 2 avoids solving large linear eigenvalue problems at all iteration steps. 
Since the energy cut-offs are determined on the fly and increased during the SCF iterations,
computational cost can be saved from that one only needs to solve small eigenvalue problems while the solution is still away from convergence.
The extra cost of this adaptive algorithm comes from calculating the error indicator, which scales as $\mathcal{O}(\Nb\Eg^{d/2}\ln\Eg)$.
Since the cost for one linear eigensolver is $\mathcal{O}(\Nb^2\Ec^{d/2})$ and $\Eg$ is proportional to $\Ec$,  we have that the additional cost for calculating the error indicator is way smaller than the cost saved from the linear eigensolvers used with small energy cut-offs.

\section{Numerical experiments}
\label{sec:numerics}

In this section, we will first test our adaptive algorithms on simple linear and nonlinear eigenvalue problems, and then show the performance on 
computing the electronic structure of typical bulk systems using the PAW method.
The approximation errors are calculated with respect to the reference solutions obtained with a sufficiently large energy cut-off.

\subsection{Tests of principle}
\label{sec:testPrinciples}

\textbf{A linear eigenvalue problem.}
Consider the 2D linear eigenvalue problems of the form \cref{eigen}, with $\Omega = [-5,5]^2$ and $V(x,y) = 1.5 (x^2 + y^2) + e^{-(x-1)^2-y^2}$.
We use a large energy cut-off $\Ec=100.0$ to obtain a sufficiently accurate solution as the reference in this example.

\begin{figure}[htb!]
	\centering 
	\includegraphics[width=0.5\textwidth]{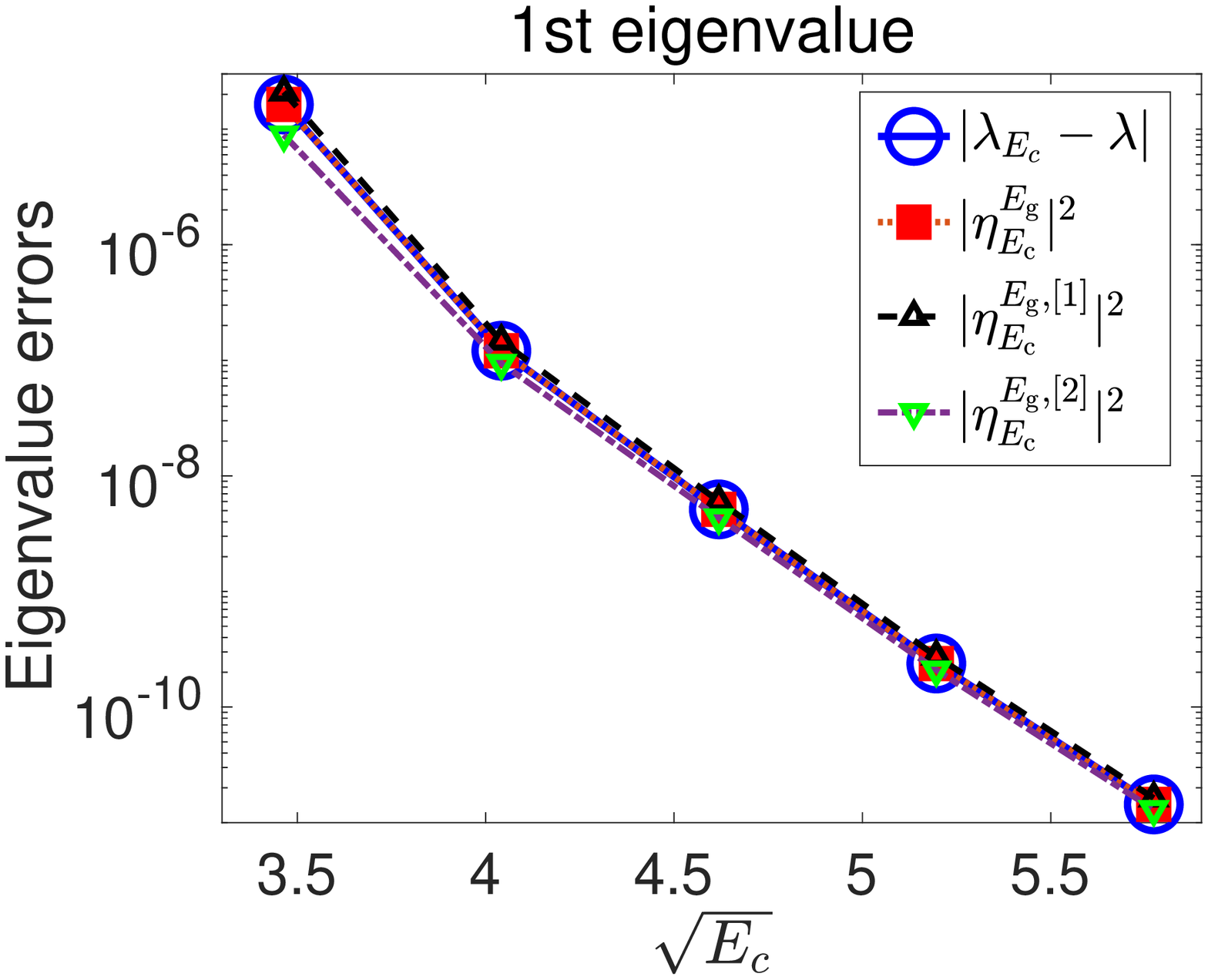}%
	\includegraphics[width=0.5\textwidth]{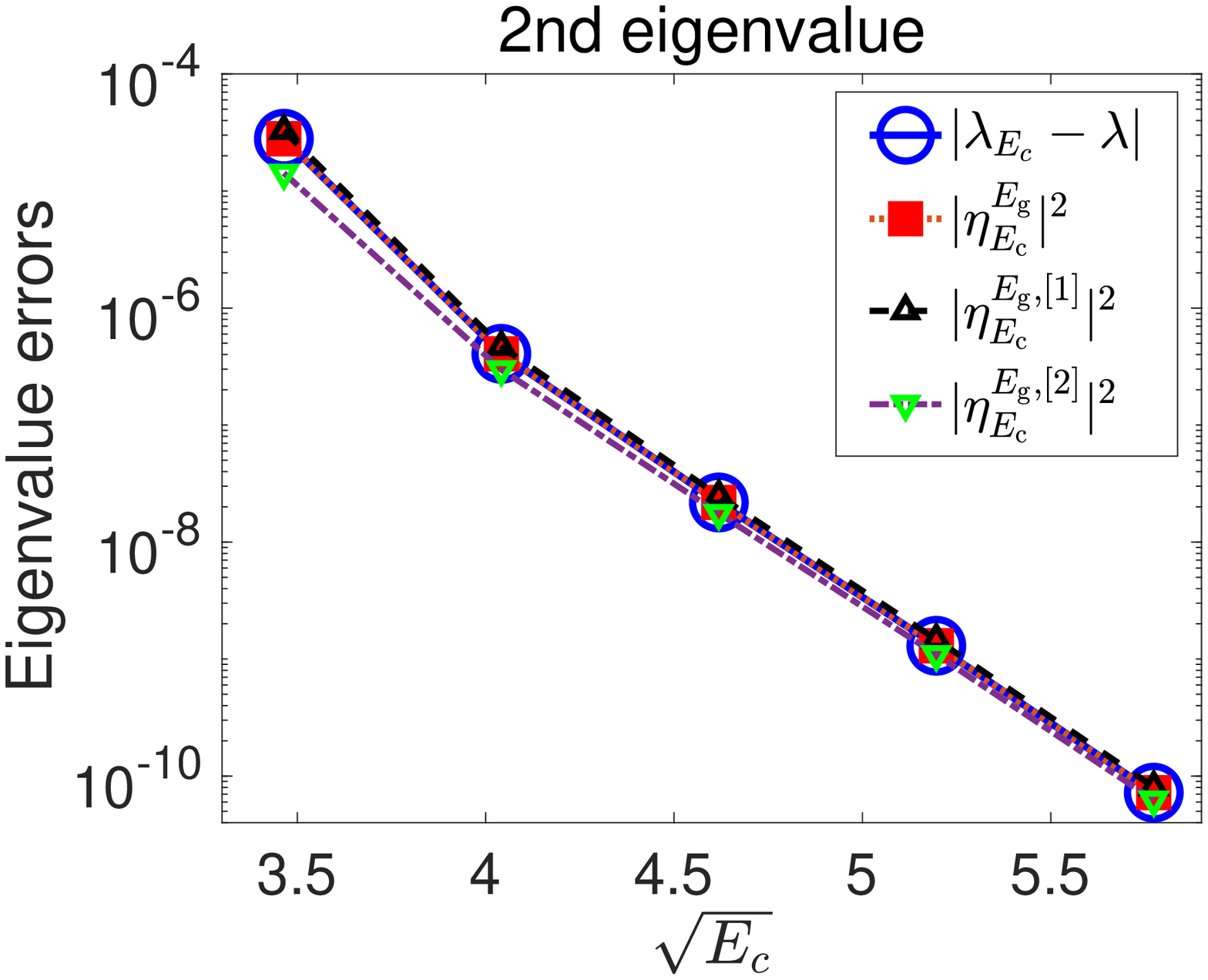}
	\caption{Linear problem: Numerical errors of the 1st and 2nd eigenvalues and the {\it a posteriori} error estimators.
		The quantity $|\eta_{E_c}^{E_g}|^2$ is defined in~\cref{err:res:local:E} and $|\eta_{E_c}^{E_g,[1]}|^2$ and $|\eta_{E_c}^{E_g,[2]}|^2$ are defined in~\cref{err:computable:Eg-k}.
	}
	\label{fig:linear:err}
\end{figure}

We first compare in \cref{fig:linear:err} the numerical errors in the lowest two eigenvalues with different {\it a posteriori} error indicators discussed in \cref{sec:posteriori,sec:evaluation}.
We observe in the plots that they match almost perfectly. 
Moreover, the errors decay exponentially with respect to the energy cut-off which suggests that a linear regression strategy could work well.
We also observe that in this case, the first order perturbation method (\cref{err:computable:Eg-k} with $k=1$) to compute the error indicator already leads to a very accurate approximation of the true error.

We further test the adaptive algorithm (Algorithm 1) on this linear problem.
\Cref{table:linear1e-3,table:linear1e-6} show the energy cut-offs obtained by using the error indicators~\cref{err:res:local:E} and \cref{err:computable:Eg-k} respectively, with Strategy A and Strategy B, and tolerances $\varepsilon = 10^{-3}$ and $10^{-6}$.
We obtain by performing a traditional convergence test that the optimal cut-offs for these two tolerances are 7.1 and 16.0 respectively. 
Algorithm~1 captures energy cut-offs respectively of 7.5 and 17.12 in a few steps, which suffice for the target accuracy without overshooting too much the optimal cut-offs.

\begin{table}[H]
	\caption{Linear problem: Error indicators and energy cut-offs obtained by Algorithm 1 with \cref{err:res:local:E}.}
	\begin{minipage}[t]{0.48\textwidth}
		\centering
		\caption*{tolerance $\varepsilon$ = 1.0e-3}
        \resizebox{\textwidth}{!}{
		\begin{tabular}{ccccc} \hline\noalign{\smallskip}
			step & $\Ec$ & $|\eta_{\Ec}^{\Eg}|^2$ & $E_{c,A}$ & $E_{c,B}$ \\
			\noalign{\smallskip}\hline\noalign{\smallskip}
			0 & 3.00 &5.121543e-01 & ------ & 7.50 \\
			\noalign{\smallskip}\hline\noalign{\smallskip}
			1 & 7.50 &5.370600e-04 &------ & ------ \\
			\hline	
        \end{tabular}}
	\end{minipage}
	\hskip 0.2cm
	\begin{minipage}[t]{0.48\textwidth}
		\centering
		\caption*{tolerance $\varepsilon$ = 1.0e-6}
        \resizebox{\textwidth}{!}{
		\begin{tabular}{ccccc} 
			\hline\noalign{\smallskip}
			step & $\Ec$ & $|\eta_{\Ec}^{\Eg}|^2$ & $E_{c,A}$ & $E_{c,B}$ \\
			\noalign{\smallskip}\hline\noalign{\smallskip}
			0 & 3.00 &5.121543e-01 & ------ & 12.00 \\
			\noalign{\smallskip}\hline\noalign{\smallskip}
			1 & 12.00 & 5.329866e-05 & 17.12 & 21.00 \\
			\noalign{\smallskip}\hline\noalign{\smallskip}
			2 &  17.12 & 4.715432e-07 & ------ & ------ \\
			\hline		
        \end{tabular}}
	\end{minipage}
	\vskip 0.2cm
	\label{table:linear1e-3}
\end{table}

\begin{table}[H]
	\caption{Linear problem: Error indicators and energy cut-offs obtained by Algorithm~1 with \cref{err:computable:Eg-k}, k = 1.}
	\begin{minipage}[t]{0.48\textwidth}
		\centering
		\caption*{tolerance $\varepsilon$ = 1.0e-3}
        \resizebox{\textwidth}{!}{
		\begin{tabular}{ccccc} 
			\hline\noalign{\smallskip}
			step & $\Ec$ & $|\eta_{\Ec}^{\Eg,[1]}|^2$ & $E_{c,A}$ & $E_{c,B}$ \\
			\noalign{\smallskip}\hline\noalign{\smallskip}
			0 & 3.00 &3.807044e-01 & ------ & 7.50 \\
			\noalign{\smallskip}\hline\noalign{\smallskip}
			1 & 7.50 &2.898487e-04  &------ & ------ \\
			\hline	
        \end{tabular}}
	\end{minipage}
	\hskip 0.2cm
	\begin{minipage}[t]{0.48\textwidth}
		\centering
		\caption*{tolerance $\varepsilon$ = 1.0e-6}
        \resizebox{\textwidth}{!}{
		\begin{tabular}{cccccc} 
			\hline\noalign{\smallskip}
			step & $\Ec$ & $|\eta_{\Ec}^{\Eg,[1]}|^2$ & $E_{c,A}$ & $E_{c,B}$ \\
			\noalign{\smallskip}\hline\noalign{\smallskip}
			0 & 3.00 & 3.807044e-01 & ------ & 12.00 \\
			\noalign{\smallskip}\hline\noalign{\smallskip}
			1 & 12.00 & 3.339172e-05 & 16.93 & 21.00 \\
			\noalign{\smallskip}\hline\noalign{\smallskip}
			2 & 16.93 & 3.395260e-07 & ------ & ------ \\
			\hline		
        \end{tabular}}
	\end{minipage}
	\vskip 0.2cm
	\label{table:linear1e-6}
\end{table}

\textbf{A nonlinear eigenvalue problem.}
Consider the following Gross--Pitaevskii equation (GPE) originated from modelling Bose--Einstein condensates~\cite{Pitaevskii}: Find $(\lambda, \varphi)\in \R\times H^1_{\#}(\Omega)$, such that $\|\varphi\|_{L^2 (\Omega)}=1$ and
\begin{flalign}
\label{eigen:gpe}
\left(-\frac{1}{2}\Delta+V(\rr) +\mu \rho(\rr) \right) \varphi(\rr) &=\lambda \varphi(\rr) 
\end{flalign}
where $\mu = 1.0$, $\Omega = [-5,5]^2$,  $V(x,y) = 10 (x^2 + y^2) + e^{-\big((x-1)^2+y^2)\big)}$, and $\rho(\rr) = |\varphi(\rr)|^2$.
Note that this equation can be viewed as a simplified version of \cref{eq:ks-standard}, where only the lowest eigenvalue is considered and the nonlinear term is much simpler.

\begin{figure}[htb]
	\centering			
	\includegraphics[width=0.48\textwidth]{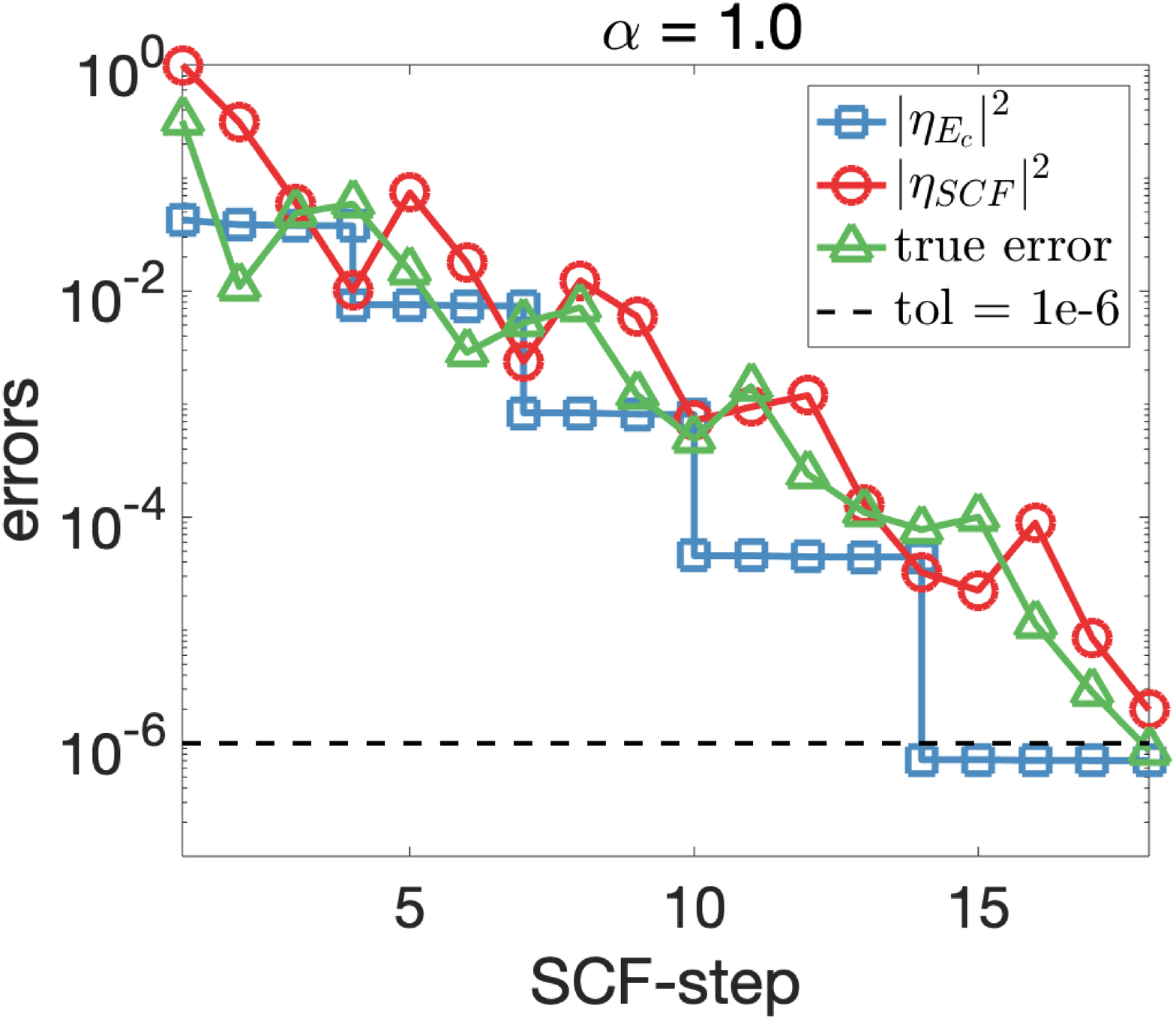}
	\includegraphics[width=0.48\textwidth]{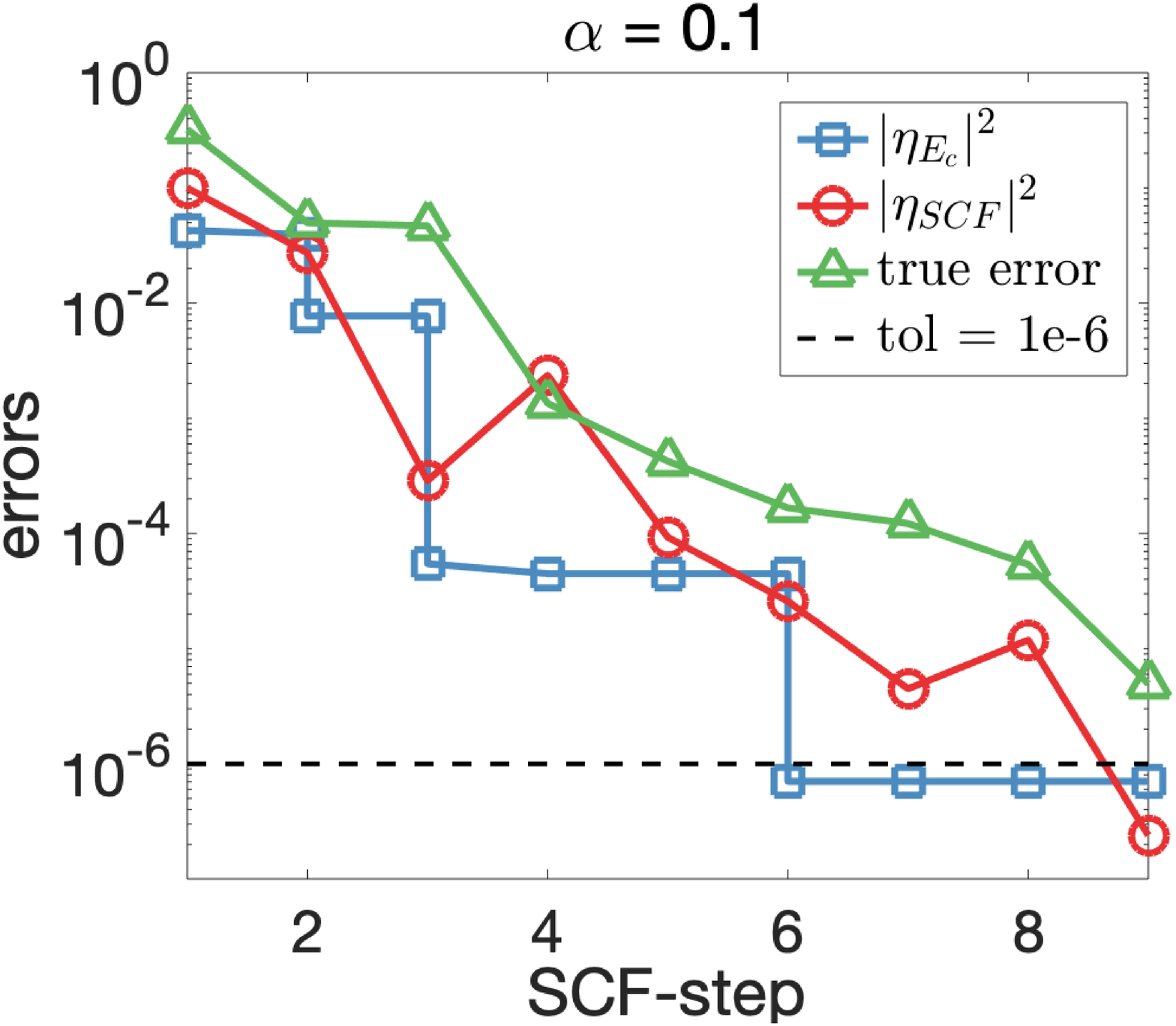}
	\caption{Nonlinear problem: Discretization, SCF and true eigenvalue errors obtained with Algorithm 2.}
	\label{fig:nonlinear}
\end{figure}

We apply Algorithm 2 to solve \cref{eigen:gpe} with tolerance $\delta=10^{-6}$.
Note that the ``best" choice of the parameter $\alpha$ (defined in \cref{err:SCF}) is not known in this example, and we perform experiments with $\alpha=1.0$ and 0.1, respectively,
to illustrate how the choice of $\alpha$ affects the number of convergence steps of our algorithm.
We show in \cref{fig:nonlinear} the discretization and SCF errors with respect to the SCF steps.
The energy cut-off is iteratively adjusted with Algorithm 2 and we observe that a good balance is achieved between the SCF and discretization errors.
The oscillation in the SCF errors arises from the simple mixing scheme used in the SCF iterations.

\subsection{Tests on PAW Kohn--Sham equations}
\label{sec:numerics_KS_PAW}

We implemented the {\it a posteriori} error indicator and the adaptive algorithm (Algorithm 2) in an in-house planewave PAW code \cite{fang2019impl}.
For the validity tests we chose two bulk systems.
The first is a diamond structure with two carbon atoms in the unit cell. This is an insulator and we used only the $\Gamma$ point for the $k$-point sampling. The number of occupied bands is $N_\uf{b}=4$.
The second system is an FCC structure with four aluminum atoms in the unit cell. 
It is a metallic system with a vanishing band gap, and we employed a $6\times 6\times 6$ Monkhorst--Pack $k$-mesh. The Brillouin zone grid is fixed over all calculations.
The first order Methfessel--Paxton smearing method \cite{methfessel1989high} with a smearing width of $0.3\,\uf{eV}$ was used and the number of occupied bands is $N_\uf{b}=7$.
For the efficiency tests on relatively large systems, we constructed a $4\times 4\times 4$
diamond supercell containing 128 carbon atoms, and the number of occupied bands is $N_\uf{b}=256$. 

In the following, we first check the accuracy of the {\it a posteriori} error indicators on the linearized Kohn--Sham equations and then test the behavior of the adaptive algorithm for solving the nonlinear equations. 
In order to take into account all the $N_\uf{b}$ occupied eigenstates, we present error results
on the band structure energy $E_\uf{bands}$.
Note that the discretization error in the band structure energy and that in the total energy decay in the same rate with respect to $E_c$ (see \cite[Theorem 4.2 and (4.90)]{cances12}).

\textbf{Accuracy of the error indicators for the linearized Kohn--Sham equations.}
We are primarily concerned with the accuracy of the error indicator for the
linearized Kohn--Sham equation since it determines the effectiveness of the
adaptive algorithm. In our tests we fixed the density as the
superposition of atomic charge densities and solved the linearized Kohn--Sham
equations for different energy cut-off values $\Ec$. The {\it a posteriori} error indicators
were calculated by using the standard and the perturbation-based methods detailed in 
\cref{sec:evaluation}.

\begin{table}
	\centering
	\caption{Comparison of estimated errors and real error in $E_\uf{bands}$
		for the linearized Kohn--Sham equation of the diamond structure.
		The $E_c$ and error data are in eV. The relative differences are
		calculated w.r.t. the real error.
		The reference solution is obtained with $\Ec=4500\,$eV.
	}
	\label{tbl:dia-err}
    \resizebox{\textwidth}{!}{
    \begin{tabular}{cccrcrcr}
        \toprule
        $\Ec$ & Real error & \multicolumn{2}{c}{$|\eta_{\Ec}^{\Eg,[1]}|^2$}
        & \multicolumn{2}{c}{$|\eta_{\Ec}^{\Eg,[2]}|^2$}
        & \multicolumn{2}{c}{$|\eta_{\Ec}^{\Eg}|^2$} \\
        \midrule
        400.0 & 2.445e-01 & 2.190e-01 &-10.44\,\% & 2.336e-01 &-4.46\,\% & 2.385e-01 &-2.45\,\%\\
        600.0 & 2.329e-02 & 2.163e-02 & -7.11\,\% & 2.256e-02 &-3.11\,\% & 2.264e-02 &-2.78\,\%\\
        800.0 & 9.093e-03 & 8.684e-03 & -4.50\,\% & 8.889e-03 &-2.25\,\% & 8.891e-03 &-2.23\,\%\\
        1000.0 & 2.690e-03 & 2.574e-03 & -4.29\,\% & 2.630e-03 &-2.21\,\% & 2.632e-03 &-2.15\,\%\\
        1200.0 & 1.684e-03 & 1.631e-03 & -3.18\,\% & 1.657e-03 &-1.65\,\% & 1.655e-03 &-1.73\,\%\\
        1400.0 & 7.283e-04 & 7.133e-04 & -2.07\,\% & 7.230e-04 &-0.73\,\% & 7.251e-04 &-0.45\,\%\\
        \bottomrule
    \end{tabular}}
\end{table}
\begin{table}
	\centering
	\caption{Comparison of estimated errors and real error in $E_\uf{bands}$
		for the linearized Kohn--Sham equation of the FCC aluminum system.
		Energy unit and relative difference calculation are the same
		as in \cref{tbl:dia-err}.
		The reference solution is obtained with $\Ec=2000\,$eV.
	}
	\label{tbl:Al-err}
    \resizebox{\textwidth}{!}{
    \begin{tabular}{cccrcrcr}
        \toprule
        $\Ec$ & Real error & \multicolumn{2}{c}{$|\eta_{\Ec}^{\Eg,[1]}|^2$}
        & \multicolumn{2}{c}{$|\eta_{\Ec}^{\Eg,[2]}|^2$}
        & \multicolumn{2}{c}{$|\eta_{\Ec}^{\Eg}|^2$} \\
        \midrule
        300.0 & 2.035e-02 & 1.805e-02 &-11.29\,\% & 1.936e-02 &-4.87\,\% & 2.000e-02 &-1.71\,\%\\
        400.0 & 4.484e-03 & 4.248e-03 & -5.27\,\% & 4.404e-03 &-1.78\,\% & 4.445e-03 &-0.88\,\%\\
        500.0 & 1.810e-03 & 1.733e-03 & -4.25\,\% & 1.790e-03 &-1.10\,\% & 1.799e-03 &-0.64\,\%\\
        600.0 & 9.420e-04 & 9.046e-04 & -3.97\,\% & 9.333e-04 &-0.92\,\% & 9.368e-04 &-0.55\,\%\\
        700.0 & 5.262e-04 & 5.032e-04 & -4.37\,\% & 5.219e-04 &-0.82\,\% & 5.240e-04 &-0.42\,\%\\
        800.0 & 3.249e-04 & 3.120e-04 & -3.96\,\% & 3.225e-04 &-0.75\,\% & 3.237e-04 &-0.37\,\%\\
        \bottomrule
    \end{tabular}}
\end{table}

The error indicators $|\eta_{\Ec}^{\Eg}|^2$,
$|\eta_{\Ec}^{\Eg,[1]}|^2$, and $|\eta_{\Ec}^{\Eg,[2]}|^2$ are compared with real
errors in \cref{tbl:dia-err,tbl:Al-err} for the diamond
structure and the aluminum systems, respectively.
It is observed from the tables that the first-order indicator $|\eta_{\Ec}^{\Eg,[1]}|^2$
already provides a very reliable approximation of the real error.
The second-order indicator $|\eta_{\Ec}^{\Eg,[2]}|^2$
exhibits a clear improvement over the first-order indicator and
derives results very close to those from the standard indicator.
And the errors estimated by the standard indicator $|\eta_{\Ec}^{\Eg}|^2$
closely match the real error.

\begin{table}
	\centering
	\caption{The CPU time cost (in seconds) of the standard and perturbation-based
		calculations of error indicators for the diamond structure.}
	\label{tbl:dia-cost}
	{ 
		\begin{tabular}{cccc}
			\toprule
			$\Ec$ & 1st-order & 2nd-order & Standard \\
			\midrule
			400.0  &  0.01  &  0.02  &  0.62 \\
			600.0  &  0.02  &  0.05  &  1.22 \\
			800.0  &  0.03  &  0.06  &  1.86 \\
			1000.0 &  0.06  &  0.11  &  3.33 \\
			1200.0 &  0.08  &  0.15  &  4.72 \\
			1400.0 &  0.10  &  0.17  &  5.87 \\
			\bottomrule
		\end{tabular}
	}
\end{table}

In terms of cost, however, the standard calculation of the error indicator
is remarkably more expensive than the perturbation-based methods,
as shown by the timings for the diamond example in \cref{tbl:dia-cost}.
As mentioned, the standard calculation requires to solve a linear system of
size $N_{\Eg}$ for each of the $N_\uf{b}$ eigenvalues.
Since the planewave discretization leads to large-scale dense matrices,
the solution of the corresponding linear systems can be very demanding.

Therefore, the perturbation-based calculation of the {\it a posteriori} error indicator
is the proper choice in practice to balance accuracy versus computational cost.
We employed the perturbation-based methods in the following tests.
It should be emphasized that the proposed error estimate is accurate both for the tested insulator and metallic systems. This high-quality {\it a posteriori} error indicator is critical to a reliable tuning of the energy cut-off.

\textbf{Adaptive solution of the Kohn--Sham equations with an unknown converged cut-off.}
We now apply Algorithm 2 to solve the nonlinear Kohn--Sham equations. 
The density mixing is performed using Pulay's DIIS method~\cite{pulay80}. The tolerance for the
discretization error in $E_\uf{bands}$ and that for the SCF error were chosen
to be $2\times 10^{-3}$\,eV for both the diamond structure and the aluminum systems. 

\begin{figure}
	\centering
	\includegraphics[width=0.48\textwidth]{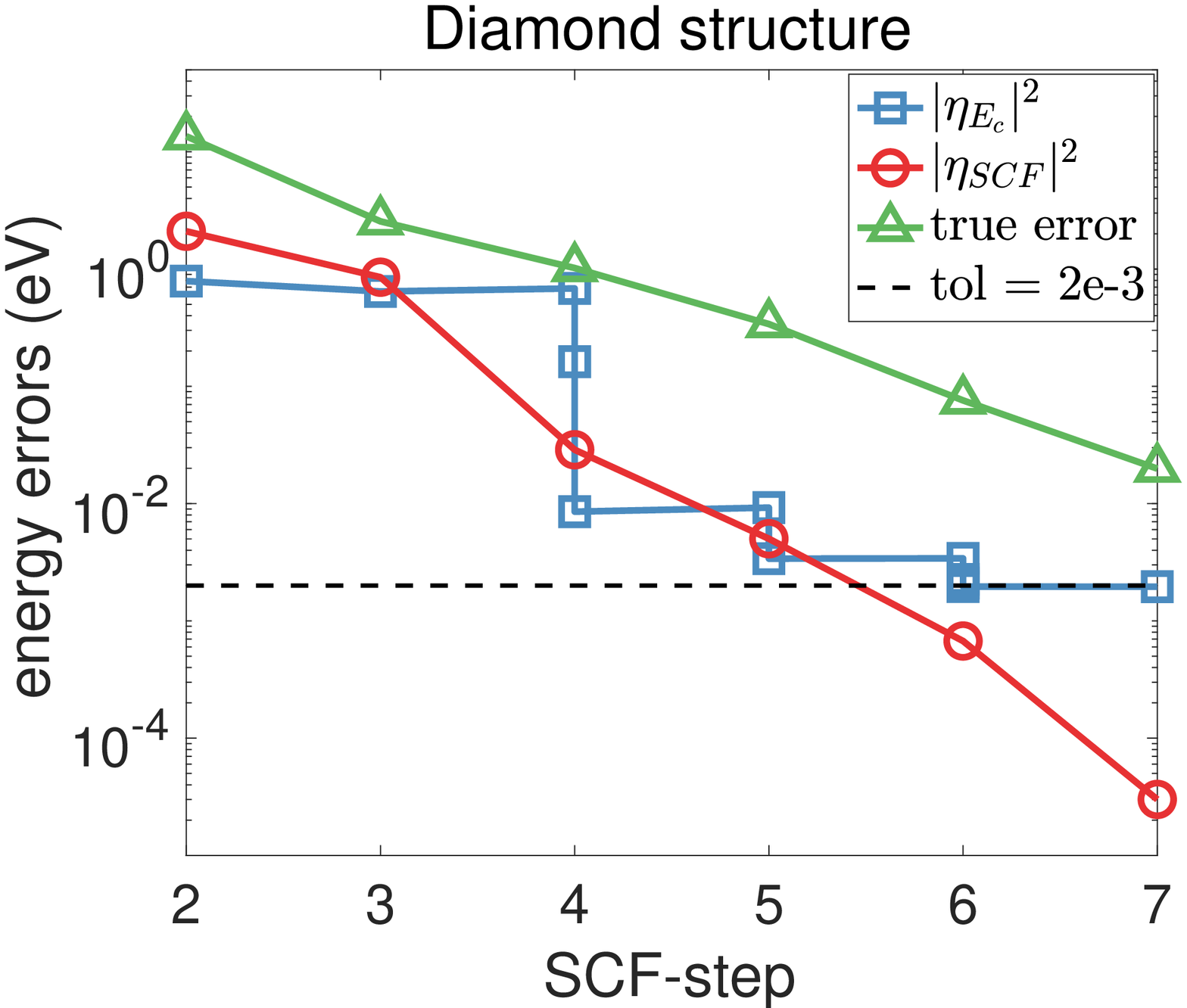}
	\includegraphics[width=0.48\textwidth]{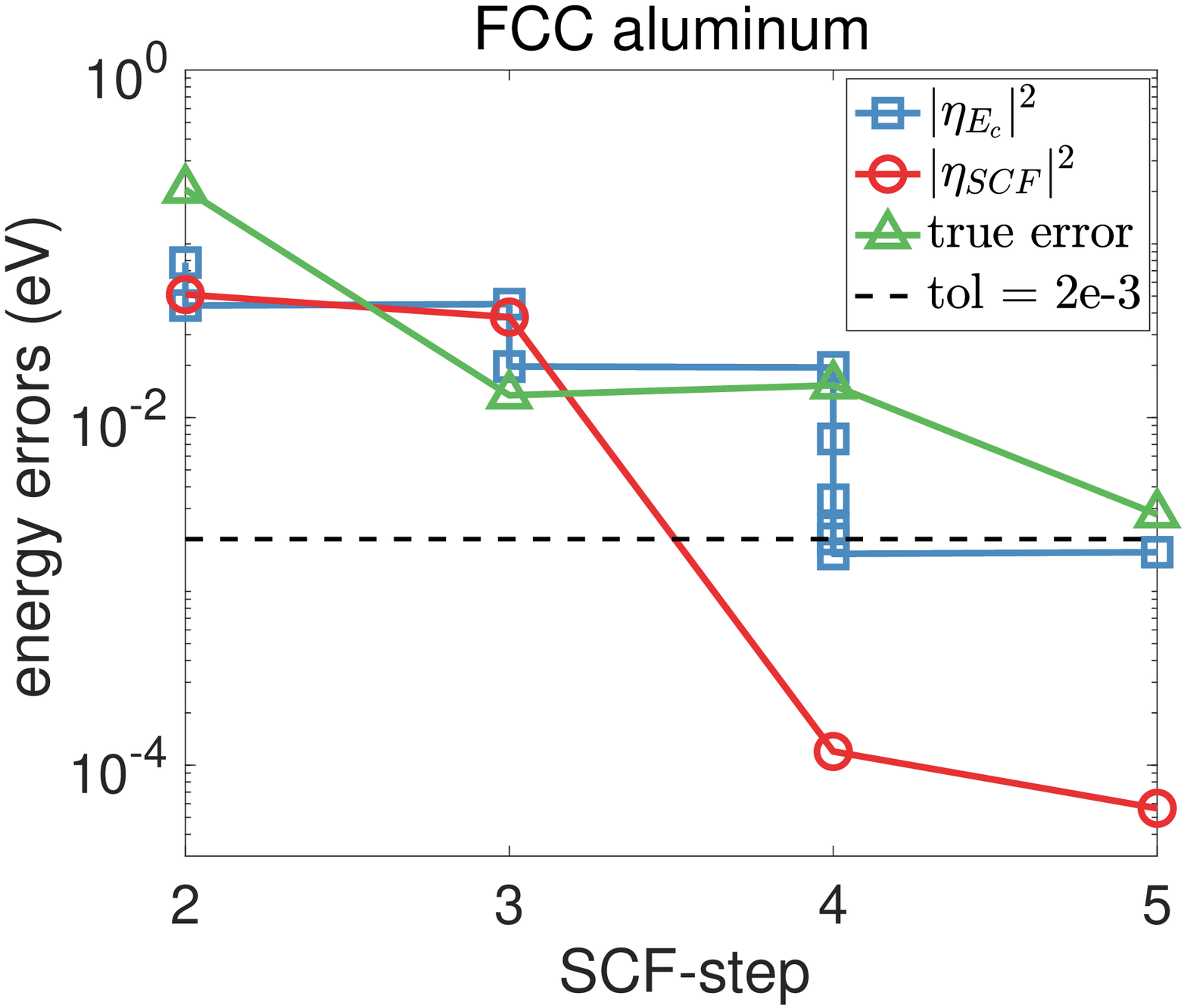}
	\caption{Discretization and SCF errors in the adaptive solution of the
    Kohn--Sham equations of the diamond structure (left) and the FCC aluminum
    system (right) using Algorithm 2.}
	\label{fig:ks-adapt}
\end{figure}

\begin{table}
	\centering
	\caption{Diamond structure: Detailed $\Ec$ and error data during the adaptive solution of the Kohn--Sham equations by Algorithm 2. The $E_c$ and errors are in eV.}
	\label{tbl:dia-adapt}
	\begin{tabular}{cccc}
		\toprule
		SCF-step & $\Ec$ & $|\eta_{\Ec}|^2$ & $|\eta_{\rm SCF}|^2$ \\
		\midrule
		2  &  300     &  7.895e-01  &  2.099 \\
		\midrule
		3  &  300     &  6.439e-01  &  8.518e-01 \\
		\midrule
		&  300     &  6.837e-01  &  \\
		4  &  400     &  1.622e-01  &  2.878e-02 \\
		&  794.42  &  8.537e-03  &  \\
		\midrule
		\multirow{2}{*}{5}
		&  794.42  &  9.242e-03  &  \multirow{2}{*}{5.018e-03} \\
		&  894.42  &  3.393e-03  &  \\
		\midrule
		\multirow{4}{*}{6}
		&  894.42  &  3.414e-03  &  \multirow{4}{*}{6.735e-04} \\
		&  994.42  &  2.260e-03  &  \\
		& 1044.42  &  2.100e-03  &  \\
		& 1094.42  &  1.960e-03  &  \\
		\midrule
		7  & 1094.42  &  1.956e-03  &  3.004e-05 \\
		\bottomrule
	\end{tabular}
\end{table}

The evolution of the discretization and SCF errors with respect to the SCF
step are shown in \cref{fig:ks-adapt} for the diamond and the aluminum systems.
For a detailed illustration, we take the diamond structure and further present detailed data on the energy cut-off and 
the two components of errors in \cref{tbl:dia-adapt}.
It is seen that a converged solution is obtained by adaptively tuning the energy cut-off depending on the discretization and SCF errors.

\textbf{Comparison with convergence-test computations and discussion.}
We have stated in the introduction that the purpose of our algorithm
is to determine the converged energy cut-off adaptively and
avoid the conventional convergence-test computations.
In this section we compare and discuss the two approaches in detail.

In a typical convergence test on the energy cut-off $\Ec$,
first, one needs to choose three parameters:
the smallest and largest cut-offs to be tested and a fixed step size of values in between.
The largest value should be greater than the converged cut-off, but the latter is unknown in advance.
So the largest cut-off needs to be set appropriately with some redundancy.
Then the Kohn--Sham equations are solved under all the chosen cut-offs,
and the converged cut-off is determined as the one beyond which
the variation in results of the concerned quantity (the energies or the forces)
is consistently smaller than the given tolerance.

In the adaptive algorithm proposed here, users need to choose the starting cut-off
and the ratio
between the tolerances for planewave discretizations and SCF iterations.
At each iteration step, the discretization error associated with $\Ec$
is estimated with the {\it a posteriori} error indicator.
Then the estimated error can be used directly to determine the computation is
converged or not. If not, the known error data are combined with
the {\it a priori} error analysis to predict the converged $\Ec$ in step sizes
not limited by the fixed cut-off step like the case of convergence tests.

It can be seen that compared with the convergence-test approach,
the proposed algorithm is based on theoretically justified bounds: the error estimates.
As for the computational cost, two factors should be taken into account here:
(i) the extra computational cost of the {\it a posteriori} error estimation
and (ii) the number of iterations used to reach the converged solution.
The time complexity for calculating the {\it a posteriori} error indicator is
$\mathcal{O}(N_\uf{b}\,N_{\Eg}\ln N_{\Eg})$
and that for the iterative diagonalization of eigenvalue problems is
$\mathcal{O}(N_\uf{b}^2\,N_{\Ec})$.
When the number of bands $N_\uf{b}$ is relatively large,
the additional cost for the error estimation is
small compared to the cost for solving eigenvalue problems.
We expect that the efficiency advantage of the adaptive algorithm
over the convergence-test approach grows with the number of bands.
The number of iterations to convergence is well controlled in the adaptive
algorithm if the tuning of $\Ec$ proceeds efficiently by
combining the {\it a posteriori} error with the {\it a priori} analysis
and using strategies stated in \cref{sec:adaptive:linear}.

In many practical simulations, 
the number of occupied bands $N_\uf{b}$ is large,
and the convergence tests are computationally demanding.
This is the case for systems with defects or vacancies, surface systems, etc.,
where large supercells must be used. This happens as well 
 for high-temperature simulations,
where the cell may be small but $N_\uf{b}$ grows rapidly with the temperature.
In the latter case, besides using a large parameter $N_\uf{b}$, the converged cutoff energy $\Ec$ is usually
significantly larger than the one needed in simulations at ambient temperature
\cite{blanchet2020}.

In the following comparison of the convergence-test approach and the adaptive algorithm,
we use a $4\times 4\times 4$ diamond supercell containing 128 carbon atoms.
The tolerance for the discretization error in $E_\uf{bands}$ and
for the SCF iteration error was set respectively to 0.128\,eV (i.e. 1\,meV/atom) and 0.1\,meV.
Note that although for simple bulk systems it is not necessary to perform convergence tests
using large supercells, we still chose the example here because the scale of the system
represents well the computational cost of systems with relatively large $N_\uf{b}$.

\begin{table}
	\centering
	\caption{Energy results (in eV/atom) and time costs (in seconds) in the
    convergence test for the Kohn--Sham equation of the diamond supercell.
    The cut-off values increase from 300\,eV to 1200\,eV with an interval of 100\,eV.}
	\label{tbl:dia-conv-100}
	\begin{tabular}{ccccccc}
		\toprule
        & \multicolumn{3}{c}{Start from scratch} 
        & \multicolumn{3}{c}{Restart from previous step} \\ \midrule
        $\Ec$ & $E_\uf{bands}/N_\uf{atom}$ & $N_\uf{SCF}$ & $t_\uf{eigen}$ & 
          $E_\uf{bands}/N_\uf{atom}$ & $N_\uf{SCF}$ & $t_\uf{eigen}$ \\ \midrule
         300 & 3.1068 & 11 &  31.23 & 3.1068 & 11 & 30.65 \\  
         400 & 3.0939 & 11 &  46.07 & 3.0932 &  6 & 26.19 \\  
         500 & 3.1027 & 11 &  61.54 & 3.1078 &  4 & 21.63 \\  
         600 & 3.0944 & 11 &  77.49 & 3.0942 &  4 & 24.93 \\  
         700 & 3.0868 & 11 &  96.73 & 3.0865 &  4 & 31.01 \\  
         800 & 3.0833 & 11 & 112.68 & 3.0830 &  4 & 32.78 \\  
         900 & 3.0820 & 11 & 139.62 & 3.0819 &  4 & 39.54 \\  
        1000 & 3.0818 & 11 & 174.29 & 3.0817 &  2 & 25.18 \\  
        1100 & 3.0816 & 11 & 193.27 & 3.0815 &  2 & 27.74 \\  
        1200 & 3.0813 & 11 & 233.94 & 3.0812 &  2 & 32.95 \\ \midrule
        Total & & &         1166.86 & & &          292.60 \\ \bottomrule
    \end{tabular}
\end{table}

\begin{table}
	\centering
	\caption{Time cost (in seconds) of the adaptive algorithm for solving the
        Kohn--Sham equation of the diamond supercell where $t_\uf{eigen}$
        represents the time for solving the eigenvalue problem and $t_\uf{err}$
        the time for calculating the {\it a posteriori} error estimate.
        The reason for the discontinuous SCF step index in the following is that,
        when the discretization error is detected to be smaller than the SCF error,
        we take the discretization error as the tolerance
        and perform more than one SCF iteration until convergence.}
	\label{tbl:dia-adapt-cost}
	\begin{tabular}{cccc}
		\toprule
        SCF-step & $\Ec$ & $t_\uf{eigen}$ & $t_\uf{err}$ \\
		\midrule
		2  &  300     &  14.33  &  1.04 \\
		\midrule                                  
		\multirow{4}{*}{6}
		   &  300     &   9.76  &  1.04 \\
		   &  400     &   9.65  &  1.46 \\
		   &  538.83  &  12.78  &  2.57 \\
		   &  805.86  &  14.59  &  2.71 \\
		\midrule
		   &  805.86  &  39.91  &  5.28 \\
		10 &  905.86  &  18.48  &  5.83 \\
		   & 1004.61  &  22.89  &  7.23 \\
		\midrule
		12 & 1004.61  &  13.93  &  -- \\
		\midrule
		Total  &      & 156.32  & 27.16 \\
		\bottomrule
	\end{tabular}
\end{table}

In the convergence test, the Kohn--Sham equations were solved under $\Ec$ values
between 300\,eV and 1200\,eV with an interval of 100\,eV.
Two subsets of tests have been carried out: in one of them, the solution of the
Kohn--Sham equation under each $\Ec$ started from scratch; in the other,
the calculation restarted from eigenfunctions under the previous cut-off.
In the adaptive algorithm, we chose the same starting cut-off $\Ec=300\,$eV
and set $\alpha=1280$ (ratio between the tolerances for SCF iterations and planewave discretizations). 
Results of these tests are given in \cref{tbl:dia-conv-100,tbl:dia-adapt-cost}.

It is observed from \cref{tbl:dia-conv-100} that in the convergence tests,
$E_\uf{bands}$ converges to 1\,meV/atom for $\Ec\ge 1000\,$eV.
This cut-off is consistent with the one (1004.61\,eV) obtained in the adaptive approach.
For time costs, it is shown in \cref{tbl:dia-conv-100} that in the convergence tests starting
from scratch, the cost of solving the Kohn--Sham equations accumulated to 1166.86\,s.
This number reduced to 292.60\,s for the convergence tests with restart,
which originates from an obvious decrease in the number of SCF iterations.
The time cost of the adaptive algorithm is composed of two parts denoted by
$t_\uf{eigen}$ for solving the eigenvalue problems and
$t_\uf{err}$ for calculating the {\it a posteriori} error.
It is seen from \cref{tbl:dia-adapt-cost} that $t_\uf{eigen}$ and $t_\uf{err}$
are 156.32\,s and 27.16\,s in total, respectively.
Compared with convergence tests from scratch, the adaptive algorithm saved
84\,\% of the CPU time. Even if the adaptive algorithm is compared with
convergence tests with restart, 37\,\% of the CPU time has been saved.

\begin{table}
	\centering
	\caption{Energy results (in eV/atom) and time costs (in seconds) in the
    convergence test for the Kohn--Sham equation of the diamond supercell.
    The cut-off values increase from 300\,eV to 1300\,eV with an interval of 200\,eV.}
	\label{tbl:dia-conv-200}
	\begin{tabular}{ccccccc}
		\toprule
        & \multicolumn{3}{c}{Start from scratch} 
        & \multicolumn{3}{c}{Restart from previous step} \\ \midrule
        $\Ec$ & $E_\uf{bands}/N_\uf{atom}$ & $N_\uf{SCF}$ & $t_\uf{eigen}$ & 
          $E_\uf{bands}/N_\uf{atom}$ & $N_\uf{SCF}$ & $t_\uf{eigen}$ \\ \midrule
         300 & 3.1068 & 11 &  31.03 & 3.1068 & 11 & 30.73 \\  
         500 & 3.1027 & 11 &  60.79 & 3.1022 &  7 & 40.25 \\  
         700 & 3.0868 & 11 &  96.99 & 3.0865 &  4 & 33.83 \\  
         900 & 3.0820 & 11 & 139.50 & 3.0819 &  4 & 45.91 \\  
        1100 & 3.0816 & 11 & 192.41 & 3.0815 &  2 & 28.93 \\  
        1300 & 3.0811 & 11 & 250.13 & 3.0809 &  4 & 69.99 \\ \midrule
        Total & & &          770.85 & & &          249.64 \\ \bottomrule
    \end{tabular}
\end{table}

Further, if one chose to perform the convergence tests with an cut-off
interval of 200\,eV, as shown by the results in \cref{tbl:dia-conv-200},
the saving in time of the adaptive algorithm compared with tests from scratch and
tests with restart decreased to 76\,\% and 27\,\%, respectively.
But at the same time, the converged cut-off has changed: energy results in the table
illustrate that $E_\uf{bands}$ converged to 1\,meV/atom for $\Ec\ge 1100\,$eV,
so the converged energy cut-off is 100\,eV over-estimated.

In summary, the comparison tests illustrate the potential efficiency advantage of
the adaptive algorithm over the convergence test approach on relatively large-scale systems. 
This kind of algorithms is expected to be further improved by integrating
new theoretical advances in the future, in particular an educated choice for the parameter $\alpha$.
In our opinion, it is worthwhile implementing these algorithms in {\it ab-initio}
software packages to stimulate the application of theoretical achievements
into practical electronic structure calculations.

\section{Conclusions and perspectives}
\label{sec:conclusion}

In this paper, we constructed an adaptive planewave method for linear eigenvalue problems and further integrated it into a self-consistent field algorithm for nonlinear eigenvalue problems.
The algorithm combines the knowledge of {\it a priori} error decay and the construction of {\it a posteriori} error indicators that is asymptotically accurate and computable.
This adaptive method not only provides good energy cut-offs in planewave methods for linear eigenvalue problems, but also shows potential for the adaptive resolution of Kohn-Sham equations in electronic structure calculations.
Specifically, in the solution of PAW Kohn--Sham equations, 
the adaptive method was shown to be built on a more sound theoretical basis and
achieve a clear saving in the time costs for relatively large systems,
in the context of convergence-test computations.

For now, we have only considered the error from the planewave discretizations. 
In real calculations though, in particular to compute physical quantities, other sources of errors come into play, such as the integration over the Brillouin zone (after Bloch's decomposition~\cite{martin05}).  Standard methods rely on equally spaced grid points and there is no reliable {\it a posteriori} error estimate for this error.
This will need to be further investigated, in order to develop efficient numerical methods for periodic systems that can estimate, control and balance the errors from planewave discretizations, SCF iterations, and Brillouin zone integrations.

\appendix

\section{Evaluating the {\it a posteriori} error indicator}
\label{sec:evaluation}

In this section, we discuss how to evaluate (or approximate) the dual norm of the residual (in \cref{error:indicator}) in an accurate and efficient way within the framework of planewave discretization. We present first a standard calculation and then a perturbation-based-method.

\subsection{A standard calculation}
\label{sec:standard_eta}

Let $A=-\Delta+V$ be the linear operator with periodic boundary condition given in~\cref{eigen}.
From \cref{coercevity} we have that $A$ is positive definite up to shifting by a positive constant, and hence $A^{1/2}$ is well defined.
From definition \cref{error:indicator}, we can obtain from a standard calculation and Cauchy-Schwarz inequality  that
\begin{multline}
\label{error:res:dnorm}
\quad \eta_{\Ec} = \sup_{v \in H} \frac{\langle \res_{\Ec}, v\rangle_{H',H} }{\|v\|_{a} }
= \sup_{v \in H} \frac{\big( A^{-1/2}\res_{\Ec}, A^{1/2} v \big)}{\sqrt{\big( A^{1/2}v, A^{1/2}v\big)} }
= \big\| A^{-1/2}\res_{\Ec} \big\|_{L^2(\Omega)}
\\[1ex]
= \big\langle \res_{\Ec} , \A^{-1}\res_{\Ec} \big\rangle_{H',H}^{\frac{1}{2}} 
= \left(\sum_{\G\in\LL'} \widehat{\res_{\Ec}}_{\G} \cdot \widehat{A^{-1}\res_{\Ec}}_{\G}\right)^{\frac{1}{2}}  ,
\quad
\end{multline}
where $\widehat{\res_{\Ec}}$ and $\widehat{A^{-1}\res_{\Ec}}$ are the vectors of Fourier coefficients of $\res_{\Ec}\in H'$ and $A^{-1}\res_{\Ec}\in H$ respectively. For $\G \in \LL'$,
\begin{align*}
\widehat{\res_{\Ec}}_{\G} = \big\langle \res_{\Ec},e_{\G} \big\rangle_{H',H}  
\qquad{\rm and}\qquad 
\widehat{A^{-1}\res_{\Ec}}_{\G} = \big(A^{-1}\res_{\Ec},e_{\G}\big) .
\end{align*}

To compute \cref{error:res:dnorm}, we shall approximate $L_{\#}^2(\Omega)$ by a finite dimensional subspace $X_{\Eg}(\Omega)$ (see definition \cref{def:spaceE})
with some $\Eg\gg\Ec$, that is, approximate the sum over $\G \in \LL'$ by finitely many terms $\frac{1}{2}|\G|^2 \leq \Eg$.
We refer to \cite[Eq. (80)-(82)]{Cances3} and \cite[Theorem 3.1]{cances12} for an analysis of the numerical integration error from the cut-off $\Eg$.
In our practical simulations, we observe that when we choose $\Eg\geq 4\Ec$, then the integration error due to the cut-off $\Eg$ is negligible compared with the discretization error with cut-off $\Ec$.

Using the orthogonality of the planewave basis functions and the fact that $(\lambda_{\Ec},\varphi_{\Ec})$ satisfies \cref{eigen:planewave}, 
the Fourier coefficients of the residual can be easily computed by
\begin{equation*}
	\widehat{\res_{\Ec}}_{\G} = a(\varphi_{\Ec}, e_{\G})-\lambda_{\Ec}(\varphi_{\Ec},e_{\G}) 
	= \big(V\varphi_{\Ec}-\Pi_{\Ec}(V\varphi_{\Ec}) , e_{\G} \big)
	\qquad \forall~{\G}\in\LL' .
\end{equation*}
More precisely, $\widehat{\res_{\Ec}}$ can be approximated within $X_{\Eg}$ by 
\begin{equation}
\label{res:q}
\widehat{\res_{\Ec}}^{\Eg}_{\G} := \left\{ \begin{array}{ll}
0 & ~~ {\rm if} ~~ \frac{1}{2}|{\G}|^2\leq\Ec
\\[1ex]
\displaystyle \big(\widehat{V\varphi_{\Ec}}\big)_{\G} = (V\varphi_{\Ec},e_{\G}) & ~~ {\rm if}~~ \Ec<\frac{1}{2}|\G|^2 \leq \Eg
\end{array}
\right. .
\end{equation}

Denote by $\widehat{\res_{\Ec}}^{\Eg} = \left(\widehat{\res_{\Ec}}_{\G}\right)_{\frac{1}{2}|{\G}|^2\leq\Eg} \in\R^{N_{\Eg}}$ 
with $N_{\Eg}$ the number of planewaves in $X_{\Eg}$.
We can approximate $\widehat{A^{-1}\res_{\Ec}}$ by
\begin{equation*}
    \widehat{A^{-1}\res_{\Ec}}^{\Eg} = \left(\widehat{A^{-1}\res_{\Ec}}_{\G}\right)_{\frac{1}{2}|{\G}|^2\leq\Eg} \in X_{\Eg}(\Omega)
\end{equation*}
by solving the linear system
\begin{equation}
\label{res:inv:q}
A_{\Eg} \cdot \widehat{A^{-1}\res_{\Ec}}^{\Eg} = \widehat{\res_{\Ec}}^{\Eg} ,
\end{equation}
where $A_{\Eg}\in\R^{N_{\Eg}\times N_{\Eg}}$ has the matrix elements $\big(A_{\Eg}\big)_{ij} = a(e_{\G_i},e_{\G_j})$.
Using \cref{error:res:dnorm,res:q,res:inv:q}, we therefore approximate the {\it a posteriori} error indicator $\eta_{\Ec}$ in practical calculations by 
\begin{equation}
\label{err:computable:Eg}
\eta_{\Ec}^{\Eg} := \left( \sum_{{\G}\in\LL',~\Ec\leq\frac{1}{2}|{\G}|^2\leq\Eg} \widehat{\res_{\Ec}}^{\Eg}_{\G} \cdot \widehat{A^{-1}\res_{\Ec}}^{\Eg}_{\G} \right)^{\frac{1}{2}}  .
\end{equation}
If the potential and eigenfunction are smooth, the approximation error $\big|\eta_{\Ec}-\eta_{\Ec}^{\Eg}\big|\ll\eta_{\Ec}$ becomes small for sufficiently large cut-off $\Eg$ \cite{Cances3}.
The error indicator $\eta_{\Ec}^{\Eg}$ is therefore asymptotically accurate and fully computable.
The main cost for computing the estimator comes from solving the linear system \cref{res:inv:q}, which costs $\mathcal{O}(N_{\Eg}\ln N_{\Eg})$.
Note that if $\Eg$ is proportional to $\Ec$, the cost for calculating $\eta_{\Ec}^{\Eg}$ is $\mathcal{O}\big(\Ec^{d/2}\ln\Ec\big)$, with $d$ the dimension of the system. 

We further introduce an error estimator with a parameter $E\in(\Ec,\Eg)$
\begin{equation}
\label{err:res:local:E}
\eta_{\Ec}^{\Eg}(E) := \left( \sum_{{\G}\in \LL',~\Ec\leq\frac{1}{2}|{\G}|^2\leq E} \widehat{\res_{\Ec}}^{\Eg}_{\G} \cdot \widehat{A^{-1}\res_{\Ec}}^{\Eg}_{\G} \right)^{1/2},
\end{equation}
which represents a relatively local error estimator in reciprocal space and is used in Strategy~B.

\begin{remark}
\label{remak:indicator_Hp}
	A natural alternative approach is to construct the {\it a posteriori} error estimate by the $H^{-1}$-norm of the residual, say $\|\res\|_{H^{-1}}$, as discussed in \cite{Cances1,Cances2,Cances:hal-02127954} for Laplace eigenvalue problems.
	We show in the next subsection that it can be viewed as a first-order perturbation-method-based calculation, and can be improved without too much cost.
\end{remark}

\subsection{A perturbation-based calculation}
\label{sec:perturbation_eta}

Solving a linear system to compute the {\it a posteriori} error estimate may be expensive when the size of the basis becomes large.
In this section, we discuss a possible workaround for the computation of the {\it a posteriori} error estimate based 
on a perturbation method. Perturbation theory~\cite{Kato1976} is a classical tool and has been widely used in electronic calculations, e.g. \cite{Baroni2001,Baroni1987,Cances4}.
In particular, \cite{cances16} provides post-processing algorithms based on a perturbation method which is related to the discussion of this section. 

In our construction, we decompose $A = A_{\Ec} + \big( V- \Pi_{\Ec} V \Pi_{\Ec}  \big)$, with  
\begin{equation}\label{A0}
A_{\Ec} = -\Delta + \Pi_{\Ec} V \Pi_{\Ec} .
\end{equation}
Using a Dyson expansion, we obtain for $k\geq 1$,
\begin{align}
\label{expansion:Ainv}
\nonumber
A^{-1} &= \sum_{j=1}^k A_{\Ec}^{-1} \Big( \big( \Pi_{\Ec} V \Pi_{\Ec} - V  \big) A_{\Ec}^{-1} \Big)^{j-1} 
+ A^{-1} \Big( \big( \Pi_{\Ec} V \Pi_{\Ec} - V \big) A_{\Ec}^{-1} \Big)^k
\\[1ex]
& =: A^{-1,[k]} + A^{-1} \Big( \big( \Pi_{\Ec} V \Pi_{\Ec} - V \big) A_{\Ec}^{-1} \Big)^k,
\end{align}
with $A^{-1,[k]} =  \sum_{j=1}^k A_{\Ec}^{-1} \Big( \big( \Pi_{\Ec} V \Pi_{\Ec} - V  \big) A_{\Ec}^{-1} \Big)^{j-1}$.
Note that the remainder term $A^{-1} \Big( \big( \Pi_{\Ec} V \Pi_{\Ec} - V \big) A_{\Ec}^{-1} \Big)^k$ will quickly become small if $\|\big( \Pi_{\Ec} V \Pi_{\Ec} - V \big) A_{\Ec}^{-1} \| \ll 1$.
We then approximate the error indicator \cref{error:res:dnorm} by
\begin{equation}
\label{err:computable:k}
\eta_{\Ec}^{[k]} := \big\langle \res_{\Ec} , \A^{-1,[k]}\res_{\Ec} \big\rangle_{H',H}^{\frac{1}{2}} ,
\end{equation}
replacing $A^{-1}$ in \cref{error:res:dnorm} by $\A^{-1,[k]}$.
In practice, we shall perform the above calculations in $X_{\Eg}(\Omega)$, and use the following {\it a posteriori} error indicator with $k =1, 2.$
\begin{equation}
\label{err:computable:Eg-k}
\eta_{\Ec}^{\Eg,[k]} := \left( \sum_{{\G}\in\LL',~\Ec\leq\frac{1}{2}|{\G}|^2\leq\Eg} \widehat{\res_{\Ec}}^{\Eg}_{\G} \cdot \widehat{A^{-1,[k]}\res_{\Ec}}^{\Eg}_{\G} \right)^{\frac{1}{2}}  .
\end{equation}
The low computational cost of this estimate comes from the following observation  \cite{Cances4}: the operator $A_{\Ec}$ being block diagonal on $X_{\Ec}$ and $X_{\Ec}^\perp$, with only the Laplace operator which is diagonal in planewaves on $X_{\Ec}^\perp$, 
and the residual~\cref{res:q} vanishing on $X_{\Ec}$, the Fourier coefficients of $A_{\Ec}^{-1}\res_{\Ec}$ can be easily evaluated by 
\begin{equation*}
	\widehat{A_{\Ec}^{-1}\res_{\Ec}}^{\Eg}_{\G} = \frac{\widehat{\res_{\Ec}}^{\Eg}_{\G}}{|\G|^2}
\end{equation*}
Therefore the estimator can be computed without solving a linear system \cref{res:inv:q}, and the computational cost of \cref{err:computable:Eg-k} for $k=1$ and $k=2$ are $\mathcal{O}\big(\Eg^{d/2}\ln\Eg\big)$.
Although the scaling is the same as that of \cref{err:computable:Eg}, but the cost has been significantly reduced since one does not need to solve a linear system. 
Note that for $k\geq 3$, the above structure for convenient evaluation will be destroyed. 

Finally, we define the corresponding error estimator with a parameter $E\in(\Ec,\Eg)$
\begin{equation}
\label{err:local:perturb}
\eta_{\Ec}^{\Eg,[k]}\big(E\big) := \left( \sum_{{\G}\in\LL',~\Ec\leq\frac{1}{2}|{\G}|^2\leq E} \widehat{\res_{\Ec}}^{\Eg}_{\G} \cdot \widehat{A^{-1,[k]}\res_{\Ec}}^{\Eg}_{\G} \right)^{\frac{1}{2}}  .
\end{equation}

\section{The PAW method}
\label{sec:PAW}

In this section, we introduce the theoretical framework of the PAW method \cite{blochl1994projector,Dup2018,fang2019impl,Kresse2},
which is one of the most important electronic structure calculation methods used nowadays.
In particular, we refer to \cite{Dup2018} and \cite{Blanc2017,Blanc2020,Dup2020} for rigorous mathematical derivations and numerical analysis for the PAW method.

The PAW Kohn--Sham model is based on a linear transformation from the smooth
pseudo (PS) wave-function $\varphi_i$ to the real all-electron (AE) wave-function $\varphi_i^\uf{AE}$ defined as
\begin{equation}\label{eq:trans}
\varphi_i^\uf{AE}(\vect{r}) = \varphi_i(\vect{r}) + \sum_{I=1}^{\Nn}
\sum_n (p_n^I,\varphi_i) \big[\psi_n^I(\vect{r}) - \phi_n^I(\vect{r})\big],
\end{equation}
where $\psi_n^I$, $\phi_n^I$, and $p_n^I$ are atomic quantities called AE partial waves,
PS partial waves, and projector functions, respectively, and for a given atomic index $I$,
the index $n$ runs over the angular momentum and an additional index labeling different partial waves for the same angular momentum.
The AE partial wave $\psi_n^I$ is constructed to be identical to the
associated PS partial wave $\phi_n^I$ outside the radius of the
core region $\Omega_I$ enclosing the atomic position. The
projector functions $p_n^I$ are compactly supported in the core region $\Omega_I$ and fulfill
\begin{equation}
\int_{\Omega_I} [ p_n^I(\vect{r}) ]^* \,\phi_m^I(\vect{r})
\,\uf{d}\vect{r} = \delta_{nm}\quad\uf{for any}~n,m.
\end{equation}

Based on the PAW transformation, one can derive consistent decomposition expressions for
the electron density, the kinetic energy of electrons, and the Hartree and the
exchange-correlation energy functionals, 
and then obtain the PAW Hamiltonian by taking variation of the total energy functional with respect to the pseudo density operator.

The AE valence density $\rho$ is decomposed as
\begin{equation}
\label{eq:paw-rho}
\rho(\vect{r}) = \tilde{\rho}(\vect{r})
+ \sum_I \big[ \rho_I^1(\vect{r}) - \tilde{\rho}_I^1(\vect{r}) \big],
\end{equation}
where $\tilde{\rho}(\vect{r})=\sum_{i=1}^{N_\uf{b}}f_i|\varphi_i(\vect{r})|^2$, and
\begin{align}
\rho_I^1(\vect{r})
&= \sum_{i=1}^{N_\uf{b}}\sum_{n,m} f_i (\varphi_i, p_n^I) (p_m^I,\varphi_i)
\,[\psi_n^I(\vect{r})]^* \,\psi_m^I(\vect{r}), \\
\tilde{\rho}_I^1(\vect{r})
&= \sum_{i=1}^{N_\uf{b}}\sum_{n,m} f_i (\varphi_i, p_n^I) (p_m^I,\varphi_i)
\,[\phi_n^I(\vect{r})]^* \,\phi_m^I(\vect{r}).
\end{align}
In the derivation of the Hartree and the exchange-correlation functionals,
the so-called pseudized core charge $\tilde{\rho}_{Zc}$,
partial electronic core charge $\tilde{\rho}_c$, and compensation charge $\hat{\rho}$
need to be further introduced. In particular,
the compensation charge $\hat{\rho} = \sum_I \hat{\rho}_I$ and
\begin{equation}\label{eq:nhat}
\hat{\rho}_I(\vect{r}) = \sum_{i=1}^{N_\uf{b}}\sum_{n,m}\sum_{L,M} f_i (\varphi_i, p_n^I) (p_m^I,\varphi_i)
\,\hat{Q}_{nm}^{LM,I}(\vect{r}),
\end{equation}
where $\hat{Q}_{nm}^{LM,I}(\vect{r})$ are confined in each augmentation region.
We refer the readers to \cite{blochl1994projector,fang2019impl,Kresse2} for
the detailed explanation of these terms.

Finally, the Kohn--Sham equation in the PAW formulation is
\begin{equation}
\label{eq:ksPAW}
H[\rho] \varphi_i = \lambda_i S\varphi_i
\qquad i = 1,\cdots,N_\uf{b}
\end{equation}
with the Hamiltonian $H[\rho]$ and overlap operator $S$ detailed below.
The PAW Hamiltonian is
\begin{equation}
\label{eq:hamil}
H[\rho] = -\frac{1}{2} \Delta + \tilde{v}_\uf{eff} + v_\uf{nl},
\end{equation}
where the local potential is given by
\begin{equation}
\tilde{v}_\uf{eff}[\tilde{\rho}, \hat{\rho}](\vect{r})
= \VH[\tilde{\rho}+\hat{\rho}+\tilde{\rho}_{Zc}](\vect{r}) +
\Vxc[\tilde{\rho}+\hat{\rho}+\tilde{\rho}_c](\vect{r}),
\end{equation}
and the non-local part is given by
\begin{equation}
\big(v_\uf{nl} \varphi_i\big)({\bf r}) = \sum_I \sum_{n,m}
D_{nm}^I (p_m^I,\varphi_i) \,p_n^I(\vect{r}),
\end{equation}
in which the non-local strength $D_{nm}^I = \hat{D}_{nm}^I+D_{nm}^{1,I}-\tilde{D}_{nm}^{1,I}$ and
\begin{align}
\hat{D}_{nm}^I
&= \sum_{L,M} \int_{\Omega_I} \tilde{v}_\uf{eff}(\vect{r})
\,\hat{Q}_{nm}^{LM,I}(\vect{r}) \, \uf{d}\vect{r},\label{eq:hatDij} \\
D_{nm}^{1,I}
&= -\frac{1}{2} \int_{\Omega_I}
[\psi_n^I(\vect{r})]^* \left[ \Delta \psi_m^I(\vect{r}) \right]
\,\uf{d}\vect{r} + \int_{\Omega_I}
[\psi_n^I(\vect{r})]^* \,\psi_m^I(\vect{r}) \,v_\uf{eff}^{1,I}(\vect{r})
\,\uf{d}\vect{r},\label{eq:Dij1} \\
\begin{split}
\tilde{D}_{nm}^{1,I}
&= -\frac{1}{2} \int_{\Omega_I}
[\phi_n^I(\vect{r})]^*
\left[ \Delta \phi_m^I(\vect{r}) \right] \,\uf{d}\vect{r}
+ \int_{\Omega_I}
[\phi_n^I(\vect{r})]^* \,\phi_m^I(\vect{r})
\,\tilde{v}_\uf{eff}^{1,I}(\vect{r}) \,\uf{d}\vect{r} \\
&\quad + \sum_{L,M} \int_{\Omega_I} \tilde{v}_\uf{eff}^{1,I}(\vect{r})
\,\hat{Q}_{nm}^{LM,I}(\vect{r}) \, \uf{d}\vect{r},
\end{split}\label{eq:tDij1}
\end{align}
with on-site potentials
\begin{align}
v_\uf{eff}^{1,I}[\rho_I^1](\vect{r})
&= \VH[\rho_I^1+\rho_{Zc}^I](\vect{r}) + \Vxc[\rho_I^1+\rho_c^I](\vect{r}), \\
\tilde{v}_\uf{eff}^{1,I}[\tilde{\rho}_I^1, \hat{\rho}_I](\vect{r})
&= \VH[\tilde{\rho}_I^1+\hat{\rho}_I+\tilde{\rho}_{Zc}^I](\vect{r}) +
\Vxc[\tilde{\rho}_I^1+\hat{\rho}_I+\tilde{\rho}_c^I](\vect{r}).
\end{align}
Finally, the overlap operator in \cref{eq:ksPAW} is given by
\begin{equation}
\label{eq:paw-overlap}
\big(S\varphi_i)(\vect{r}) = \varphi_i(\vect{r}) + \sum_I \sum_{n,m}
q_{nm}^I ( p_m^I, \varphi_i) \,p_n^I(\vect{r}),
\end{equation}
where
\begin{equation}
\label{eq:qij}
q_{nm}^I = \int_{\Omega_I} \Big\{ [\psi_n^I(\vect{r})]^* \,\psi_m^I(\vect{r})
- [\phi_n^I(\vect{r})]^* \,\phi_m^I(\vect{r})
\Big\} \,\uf{d}\vect{r}.
\end{equation}

When replacing the standard Kohn--Sham equation \cref{eq:ks-standard} with the PAW formulation \cref{eq:ksPAW}, the framework of the adaptive algorithm designed in this paper stays the same, though a lot of attention needs to be paid for numerical implementations of the above terms.

\section*{Acknowledgement}
B. Liu and H. Chen's work was supported by the National Natural Science Foundation of China under grant 11971066.
G. Dusson's work was partially supported by the French ”Investissements d'Avenir” program, project ISITE-BFC (contract ANR-15-IDEX-0003).
J. Fang's work was supported by the National Natural Science Foundation of China under grant 11701037.
X. Gao's work was supported by the Science Challenge Project under Grant TZ2018002.


\end{document}